\documentclass[useAMS,usenatbib]{mn2e}

\usepackage{Times}

\usepackage{psfrag}
\usepackage{pspicture}
\usepackage{graphicx}

\title[The Nature of H$\alpha$ Emitters at z=0.84]{ HiZELS: a high redshift survey of H$\alpha$ emitters. II: the nature of star-forming galaxies at z=0.84 \thanks{Based on observations obtained with the Wide Field CAMera (WFCAM) on the United Kingdom Infrared Telescope (UKIRT)} }
\author[D. Sobral et al.]{D. Sobral$^{1}$\thanks{E-mail: drss@roe.ac.uk}, P. N. Best$^{1}$, J. E. Geach$^{2}$, Ian Smail$^{2}$, J. Kurk$^{3}$, M. Cirasuolo$^{1}$, M. Casali$^{4}$,
\newauthor R. J. Ivison$^{1,5}$, K. Coppin$^{2}$ \& G. B. Dalton$^{6,7}$ \\
$^{1}$SUPA, Institute for Astronomy, Royal Observatory of Edinburgh, Blackford Hill, Edinburgh, EH9 3HJ, UK\\
$^{2}$Institute of Computational Cosmology, Durham University, South Road, Durham, DH1 3LE, UK\\
$^{3}$Max-Planck-Institut f{\"u}r Astronomie, K{\"o}nigstuhl, 17 D-69117, Heidelberg, Germany\\
$^{4}$European Southern Observatory, Karl-Schwarzschild-Strasse 2, D-85738 Garching, Germany\\
$^{5}$Astronomy Technology Centre, Royal Observatory of Edinburgh, Blackford Hill, Edinburgh, EH9 3HJ, UK\\
$^{6}$Astrophysics, Department of Physics, Keble Road, Oxford, OX1 3RH, UK\\
$^{7}$Space Science and Technology, Rutherford Appleton Laboratory, HSIC, Didcot, OX11 0QX, UK}
\begin{document}

\date{Accepted 2009 May 22. Received 2009 May 11; in original form 2009 January 25}

\pagerange{\pageref{firstpage}--\pageref{lastpage}} \pubyear{2009}

\maketitle

\label{firstpage}

\begin{abstract}
New results from a large survey of H$\alpha$ emission-line galaxies at $z=0.84$ using the Wide Field CAMera on UKIRT and a custom narrow-band filter in the J band are presented as part of the HiZELS survey. The deep narrow-band images reach an effective flux limit of F$_{{\rm H}\alpha}\sim10^{-16}$\,erg\,s$^{-1}$\,cm$^{-2}$ in a co-moving volume of $1.8 \times 10^5$\,Mpc$^3$, resulting in the largest and deepest survey of its kind ever done at $z\sim1$. There are 1517 potential line emitters detected across $\sim1.4$ square degrees (in two fields: COSMOS and UKIDSS UDS), of which 743 are selected as H$\alpha$ emitters, based on their photometric and spectroscopic redshifts. These are then used to calculate the H$\alpha$ luminosity function, which is well-fitted by a Schechter function with $L^*=10^{42.26 \pm 0.05}$\,erg\,s$^{-1}$, $\phi^*$=$10^{-1.92 \pm 0.10}$ Mpc$^{-3}$ and $\alpha=-1.65\pm0.15$, and are used to estimate the volume average star formation rate at $z=0.845$, $\rho_{SFR}$: $0.15\pm0.01$\,M$_{\odot}$\,yr$^{-1}$\,Mpc$^{-3}$ (corrected for 15\% AGN contamination and integrated down to 2.5 M$_{\odot}$yr$^{-1}$). These results robustly confirm a strong evolution of $\rho_{SFR}$ from the present day out to $z\sim1$ and then flattening to $z\sim2$ using a single star-formation indicator: H$\alpha$ luminosity. Out to $z\sim 1$, both the characteristic luminosity and space density of the H$\alpha$ emitters increase significantly; at higher redshifts, $L^*$ continues to increase, but $\phi^*$ decreases. The $z=0.84$ H$\alpha$ emitters are mostly disk galaxies ($82\pm3$\%), while $28\pm4$\% of the sample show signs of merger activity; mergers account for $\sim 20$\% of the total integrated $\rho_{SFR}$ at this redshift. Irregulars and mergers dominate the H$\alpha$ luminosity function above $L^*$, while disks are dominant at fainter luminosities. These results demonstrate that it is the evolution of ``normal`` disk galaxies that drives the strong increase in the star formation rate density from the current epoch to $z \sim 1$, although the continued strong evolution of $L^*$ beyond $z=1$ suggests an increasing importance of merger activity at higher redshifts.

\end{abstract}

\begin{keywords}
galaxies: high-redshift, galaxies: luminosity function, cosmology: observations, galaxies: evolution.
\end{keywords}

\section{Introduction}

Understanding the basic features of galaxy formation and evolution requires unveiling the volume-averaged star formation rate as a function of epoch, $\rho_{SFR}$, its distribution function within the galaxy population, and its variation with environment. At the present, surveys of the star-formation rate density suggest a rise of $(1+z)^4$ out to at least $z\sim1$ \citep[e.g.][]{1996ApJ...460L...1L,con,1998AAS...192.5104H,2000ApJ...544..641H,2004ApJ...615..209H,2006ApJ...651..142H} indicating that most of the stars in galaxies today formed at $z>1$.

In the local Universe,  studies have demonstrated that star formation is strongly dependent on the environment. While clusters of galaxies seem to be primarily populated by passively-evolving galaxies, star-forming galaxies are mainly found in  less dense environments \citep[e.g.][]{2002MNRAS.334..673L}. Star-forming galaxies have also been found to have lower masses than passive galaxies  \citep{1996AJ....112..839C}. How do these environmental and mass dependencies change with cosmic time? When did they start to be noticeable, and how do they affect the evolution of galaxies, clusters and the star formation rate density of the Universe as a whole? How much of the evolution of the cosmic star formation rate density is associated with the evolution of star-forming galaxies and how much is driven by galaxy merger activity?  In order to properly answer such questions it is mandatory to conduct observational surveys at high redshift, which can then be used to test theoretical models of galaxy evolution. These can be performed with a large variety of techniques and instrumentation.

There are many different star formation indicators which have been widely used both to select and study star-forming galaxies, such as the  Ultra-Violet (UV) light emitted by young stars, the energy that is absorbed and then re-emitted in the form of emission lines (such as H$\alpha$ or [O{\sc ii}]\,3727), or in the sub-mm and infra-red, or even radio emission from supernovae. Ideally, the use of different star formation tracers would provide consistent answers to most of the questions that remain unanswered. Unfortunately, studies with different star formation tracers suffer different biases and selection effects, which contribute to considerable discrepancies. These problems are also amplified by the effects of cosmic variance in most of the current samples. Another issue is the difficulty in correcting for extinction, especially for UV and optical wavelengths, which can lead to large systematic uncertainties in the star formation densities derived from measurements in these wavebands.

No single indicator provides a perfect view of the evolution of the star-formation rate density \citep{kewe}, but using different tracers is not the solution either. Ideally, one needs to choose a star-formation indicator that can be applied from the present to high redshift, is relatively immune to dust extinction, and which has sufficient sensitivity to ensure that the derived star-formation rate distribution does not require large extrapolations for faint sources below the sensitivity limit. The H$\alpha$ luminosity is arguably the best candidate to achieve these goals and large-area surveys with a sensitivity of a few solar masses per year can now be done with current instrumentation out to $z\sim2$. This contrasts with the equivalent star formation rate limits for dust-independent tracers such as radio, far-infrared or sub-millimeter observations which vary from $\sim10$--100\,M$_{\odot}$\,yr$^{-1}$ at the same redshifts for large area surveys.

Large H$\alpha$ surveys have been carried out since the 1970's \citep[e.g.][]{cohen76, kenkent, Roma90, Gava91,galo,yan,alpha1,doherty}. With the rise of large-format imaging cameras in the optical and in the near-infrared, narrow-band filters can now be used to undertake deep surveys for emission-line objects in large volumes which cover the great majority of the cosmic history, out to $z\sim2.5$ using H$\alpha$ \citep[e.g.][]{2008ApJ...677..169V, 2008arXiv0805.2861G}. The surveys identify sources on the basis of the strength of their emission line and thus represent roughly a star-formation rate selected sample. Hence, by using a set of narrow-band filters, it is possible to apply a single technique to target H$\alpha$ emitters across a wide range of redshifts, gathering representative samples at each epoch with a uniform selection. Those samples are powerful tools for tracing the evolution in the star-formation rate density across the expected peak of star-formation in the history of the Universe, which is one of the main aims of the High-$z$ Emission Line Survey (HiZELS).

HiZELS is a panoramic extragalactic survey using the wide area coverage of the Wide Field CAMera (WFCAM, Casali et al.\ 2007) instrument on the 3.8-m UK InfraRed Telescope (UKIRT). The survey utilizes a set of existing and custom-made narrow-band filters in the $J$, $H$ and $K$ bands to detect emission line galaxies over $\sim 5$ square degrees of extragalactic sky. The H$_2$S(1) narrow-band filter is being used to target H$\alpha$ emitting galaxies at $z=2.23$, which are predominantly star-forming systems (Geach et al.\ 2008). In addition, narrow-band filters in the $J$ and $H$ bands (hereafter NB$_J$ and NB$_H$) have been custom-designed to target the [O{\sc ii}]\,3727 and [O{\sc iii}]\,5007 emission lines in galaxies at the same redshift as the H$_2$S(1) H$\alpha$ survey. Together, the three sets of filters help to detail the properties of the line emitters at $z=2.23$, while the NB$_J$ and NB$_H$ filters deliver identically-selected H$\alpha$ samples at $z=0.84$ and 1.47 respectively.

%
%
\begin{table*}
 \centering
  \caption{Observation log for the NB$_J$ and $J$ observations of the COSMOS  and UKIDSS UDS fields.}
  \begin{tabular}{@{}cccccccccc@{}}
  \hline
   Field & Filter & R.A. & Dec. & Int.\ time & FHWM  & Dates  & $m_{lim}$  \\
        & & {(J2000)} &(J2000) & (ks) & ($''$) & (2007) & (3$\sigma$) \\
 \hline
   \noalign{\smallskip}
 COSMOS 1 & NB$_J$ & 10\,00\,00 & +02\,10\,30 & 19.7 & 1.0 & 14--16 Jan & 21.7  \\
 COSMOS 2 & NB$_J$& 10\,00\,52 & +02\,10\,30 & 21.6 & 1.0 & 13, 14 Jan& 21.6 \\
 COSMOS 3 & NB$_J$& 10\,00\,00 & +02\,23\,44 & 19.0 & 0.9 & 15--17 Jan& 21.7 \\
 COSMOS 4 & NB$_J$& 10\,00\,53 & +02\,23\,44 & 17.2 & 0.9 & 15, 17 Jan, 13--16 Feb& 21.6 \\
 COSMOS 1 & $J$ & 10\,00\,00 & +02\,10\,30 & 5.7 & 0.9 & 14--16 Jan& 22.8  \\
 COSMOS 2 & $J$ & 10\,00\,52 & +02\,10\,30 & 6.9 & 1.0 & 13, 14 Jan& 22.8  \\
 COSMOS 3 & $J$ & 10\,00\,00 & +02\,23\,44 & 5.7 & 0.9 & 15--17 Jan& 22.8   \\
 COSMOS 4 & $J$ & 10\,00\,53 & +02\,23\,44 & 5.1 & 0.9 & 15, 17 Jan, 13--16 Feb& 22.6   \\
 UDS NE & NB$_J$ & 02\,18\,29 & $-$04\,52\,20 & 21.2 & 1.0 & 18, 20 Oct & 21.6 \\
 UDS NW & NB$_J$ & 02\,17\,36 & $-$04\,52\,20 & 22.6 & 1.0 & 19, 21 Oct & 21.7\\
 UDS SE & NB$_J$ & 02\,18\,29 & $-$05\,05\,53 & 20.0 & 1.0 & 19 Oct & 21.5  \\
 UDS SW & NB$_J$ &  02\,17\,38 & $-$05\,05\,34 & 22.5 & 1.0 & 20, 21 Oct & 21.6  \\
 \hline
\end{tabular}
\label{obs}
\end{table*}

This paper presents deep narrow-band imaging using the NB$_J$ filter at $\lambda = 1.211\umu$m, as part of HiZELS, over $\sim 0.7\deg^2$ of the in the SXDF Subaru-XMM--UKIDSS Ultra Deep Survey \citep{2007MNRAS.379.1599L} field (UDS), and $\sim 0.8\deg^2$ in the Cosmological Evolution Survey \citep{2007ApJS..172....1S,2007ApJS..172..196K} field (COSMOS). This corresponds to an area coverage which is $\sim 8$ times larger and twice the depth of the survey by \cite{2008ApJ...677..169V}, the best previous emission-line survey at $z\sim1$.

The paper is organised in the following way. \S2 outlines the details of the observations, and describes the data reduction, photometric calibration, source extraction and survey limits. \S3 presents the narrowband selection criteria, the final sample and the photometric and spectroscopic redshift analysis. Results are presented in \S4: the H$\alpha$ luminosity function, the star-formation rate density, the morphologies of the H$\alpha$ emitters and their relation and contribution to the H$\alpha$ luminosity function. Finally, \S5 outlines the conclusions. An H$_0=70$\,km\,s$^{-1}$\,Mpc$^{-1}$, $\Omega_M=0.3$ and $\Omega_{\Lambda}=0.7$ cosmology is used  and all magnitudes are in the Vega system, except if noted otherwise.

\section{OBSERVATIONS and DATA REDUCTION}

Both fields were observed with WFCAM on UKIRT using a set of custom narrow-band J filters ($\lambda = 1.211\umu$m, $\delta\lambda = 0.014\umu$m), during 2007 January 13--17 and February 13,14 \& 16 for the COSMOS field and October 19--24 for UKIDSS UDS. WFCAM's standard ``paw-print'' configuration of four $2048\times2048$ $0.4''$\,pixel$^{-1}$ detectors offset by $\sim20'$ can be macrostepped four times to cover a contiguous region of $\sim55'\times55'$ \citep{2007A&A...467..777C}. For each field (COSMOS and UDS) 4 paw-prints, or $\sim0.8 \deg^2$ were mapped with narrowband exposures of $\sim21$\,ks\,pixel$^{-1}$. The seeing varied between $0.8$--$1.0''$ during the observing nights, and conditions were photometric. The NDR (Non Destructive Read) mode was used for all narrow-band observations to minimise the effects of cosmic rays in long exposures. The observations were obtained following a 14-point jitter pattern for UDS (to match the UKIDSS observing strategy) and a 9-point jitter sequence for COSMOS. Broad-band $J$-band observations were also obtained in the COSMOS field, but were not necessary for the UDS due to the availability of UKIDSS data. A summary of the observations is given in Table~\ref{obs}.

\subsection{Data reduction}

In order to accomplish the survey goals, a dedicated pipeline has been developed: PfHiZELS. This was done following the method presented in \cite{2008arXiv0805.2861G}, improving several steps and producing a user-friendly, completely automated pipeline. It consists of a set of {\sc python} scripts which implement a complete reduction of WFCAM data, taking advantage of several publicly available packages, as detailed below. Small samples of data were also reduced by hand using {\sc iraf} (including {\sc dimsum} and {\sc xdimsum}), which allowed a comparison with the automatic data reduction and showed that the results are comparable.

Dark frames were median combined for each observing night and the individual science frames were then dark subtracted. Next, frames were median combined in each jitter sequence of 14 (UDS) or 9 (COSMOS) without any offset, to produce a rough first pass flat. The latter was then used to produce a badpixel mask for each chip by flagging pixels which deviate by more than $3\sigma$ from the median value. Each frame was then flattened, and an individual badpixel mask (for each frame) was produced.

In order to produce better flats, {\sc SExtractor} \citep{1996A&AS..117..393B} was used to identify objects on the flattened images, and source masks were created and used to produce second-pass flat fields. These were then used to create the final flattened frames.

A world coordinate system was fitted to each frame by querying the USNO A2.0 catalog. This fitted, on average, $\sim 100$ objects per frame and rotated images appropriately for each WFCAM chip. Frames were then de-jittered and co-added with {\sc SWarp} (Bertin 1998), which performs a background mesh-based sky subtraction optimized for narrow-band data. It should be pointed out that WFCAM frames suffer from significant cross-talk artifacts, which manifest themselves as toroidal features at regular (every 128) pixel intervals from sources, in the read-out direction. Furthermore, as these are linked to a ``physical'' location, they can only be removed from the source catalog (see \S\ref{extraction}).

\subsection{Photometry calibration}

Narrow- and broad-band images were photometrically calibrated (independently) by matching $\sim70$ stars with $m_J=11$--16 per frame from the 2MASS All-Sky catalog of Point Sources \citep{2003tmc..book.....C} which are unsaturated in narrow-band frames. Based on the photometric calibration, the absolute calibration is expected to be accurate to $<3\%$ when compared to 2MASS. We have also double-checked the offset between zero-points by comparing them to the expected values, roughly given by $-2.5\log\frac{\delta \lambda_J\times \delta t_J}{\delta \lambda_{NB_J}\times \delta t_{NB_J}}$, where $\delta \lambda$ is the filter width and $\delta t$ the exposure time per frame. For convenience, after this step, narrow and broad-band images were normalised to give them the same zero-point.

\subsection{Source Extraction and Survey Limits}
\label{extraction}

The survey is made up of a mosaic of eight WFCAM pointings (4 each for UDS and COSMOS), i.e., $8\times4\times13.7'\times13.7'$ tiles. The UKIDSS UDS $J$-band image does not overlap entirely with the full narrow-band image, leading to a total overlapping area of  $0.70 \deg^2$, while for COSMOS the overlap is $0.78 \deg^2$.

Sources were extracted using {\sc SExtractor} \citep{1996A&AS..117..393B}. Optimal parameters were found by running a large number of different extractions which converged to a set of parameters that allowed the extraction of all obvious sources down to the 3-$\sigma$ limit in each frame and minimized the extraction of noise/artifact features. The extraction included an optimized sky subtraction, and fixed photometry apertures of $3''$ (diameter) were used. Several tests were done using {\sc SExtractor} in dual mode (using frames in one band to detect the sources and measuring on the other band) and single mode. The first option was used for COSMOS (where data were taken in both bands and so the frames were extremely well registered), while for UDS (where UKIDSS data was used for the $J$ band) the extraction was done on both bands independently, followed by a match using a simple $2.0''$ criteria. In 11 cases there was more than one match; for these, a careful local astrometry solution was calculated based on nearby sources clearly identified in both bands and the correct source was then clear in all cases (the correct match was always the original closest). Furthermore, tests were run on COSMOS frames to show that both options produce comparable samples, and so this is not likely to produce any significant difference between the extraction for UDS and COSMOS. Narrowband sources with no clear $J$ band detection were retained but there are few of these as the $J$ coverage is significantly deeper than the NB$_J$. As the observations have slightly different FHWM, different total exposure times, and because each WFCAM chip has slightly different properties, objects were extracted down to each chip's limit, which was then confirmed using Monte Carlo simulations (see \S\ref{compl}).

%
%
\begin{figure}
\centering
\includegraphics[width=8.2cm]{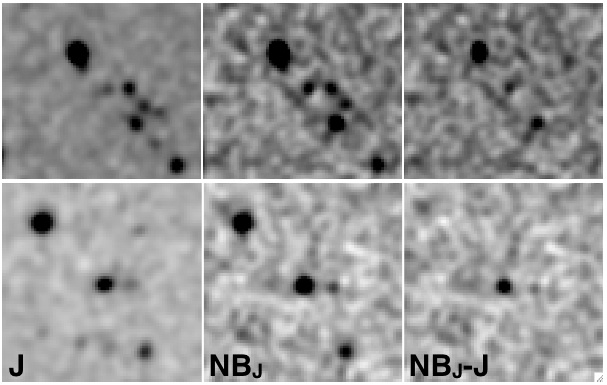}
\caption[Excess Sources in COSMOS]{Two narrow-band excess objects selected from the NB$_J$ imaging with different colour significances (upper panels), and a strong isolated emitter (lower panels). Images are $\sim20''\times20''$. \label{excess}}
\end{figure}

In order to clean spurious sources from the catalogue (essential to remove cross-talk artifacts), the frames were visually inspected showing that sources brighter than $\sim12$\,mag are surrounded by large numbers of artifacts detected within ``bright halos'', as well as cross-talk, while fainter sources (up to 16\,mag) show only cross-talk features. Sources fainter than 16\,mag do not produce any detectable cross-talk at the typical depth of the observations. Cross-talk sources and detections in the halo regions were removed from the catalogue separately for each frame, which greatly simplifies their identification (Geach et al.\ 2008). 

Catalogue sources detected in regions with less than 85\% of the total integration time were removed; which still assures a complete overlap between frames. When a source was catalogued in more than one final image, the catalogue entry with a higher exposure time was selected.

The average 3-$\sigma$ depth of the entire set of NB$_J$ frames is 21.6\,mag, with $J$ depths being $J=23.4$\,mag (UKIDSS UDS DR3) and $J=22.7$\,mag (COSMOS). The narrow-band imaging detects a total of 21773 objects in COSMOS across $0.76\deg^2$ and 15449 in UDS across $0.68 \deg^2$. These areas include the removal of regions in which cross-talk and other artifacts caused by bright objects are located.

\section{SELECTION}

\subsection{Narrowband excess selection}

Emission line systems are initially selected according to the significance of their $(J-$\,NB$_J)$ colour, as they will have $(J-$\,NB$_J)>0$. However, true emitters need to be distinguished from those with positive colours due to scatter in the magnitude measurements and this is done by quantifying the significance of the narrowband excess. The parameter $\Sigma$ quantifies the excess compared to the random scatter expected for a source with zero colour \citep{1995MNRAS.273..513B}. In other words, a source can be considered to have a genuine narrow-band excess if:

\begin{equation}
   c_{NB_J}-c_{J}>\Sigma\delta
\end{equation}
where $c_{{\rm NB}_J}$ and $c_{J}$ are the counts for the NB$_J$ and $J$ bands, respectively, and $\delta$ is the combined photometric error:

\begin{equation}
  \delta=\sqrt{\pi r^2(\sigma_{{\rm NB}_J}^2+\sigma_{J}^2)}.
\end{equation}
The approach is similar to a standard signal-to-noise selection. Colour and $\Sigma$ significances are related by

\begin{equation}
   J-{{\rm NB}_J}=-2.5\log(1-\Sigma\delta10^{-0.4(ZP-{{\rm NB}_J})}),
\end{equation}
where $ZP$ is the zeropoint of both frames, assuming that the sky variation is the dominant feature contributing to the errors in photometry. This has been tested by measuring counts in randomly placed apertures ($\sim 10000$ for each frame), confirming the hypothesis.

%
%
\begin{figure*}
\begin{minipage}[b]{0.48\linewidth} 
\centering
\includegraphics[width=8.2cm]{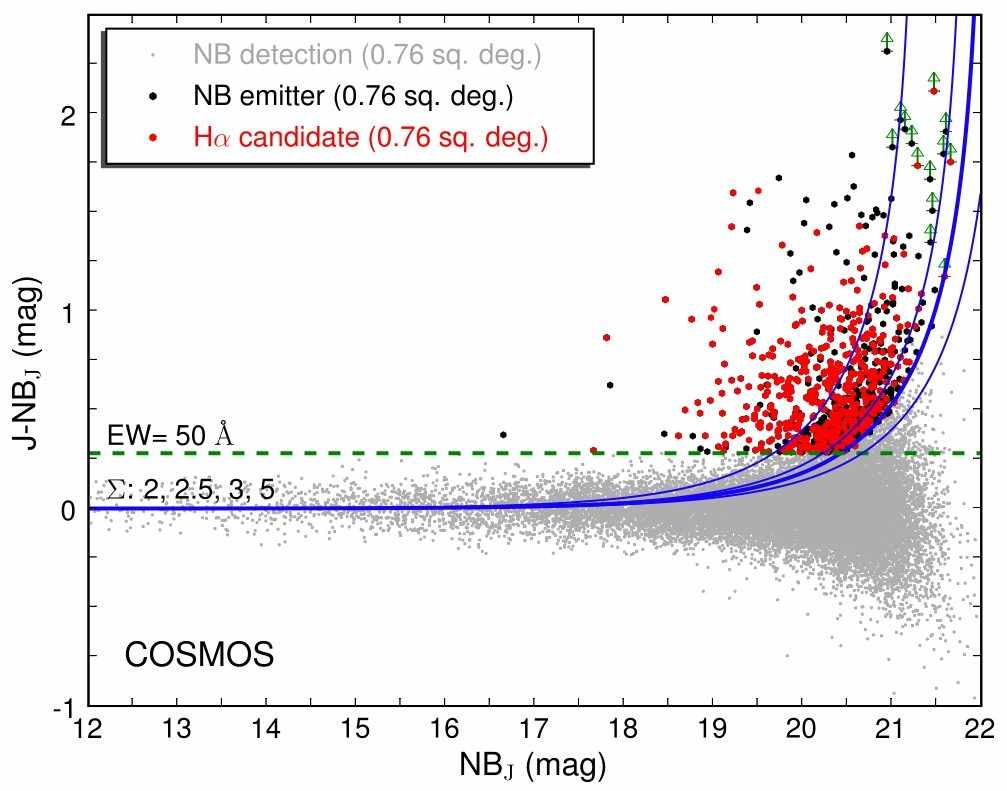}
\end{minipage}
\hspace{0.1cm} 
\begin{minipage}[b]{0.48\linewidth}
\centering
\includegraphics[width=8.2cm]{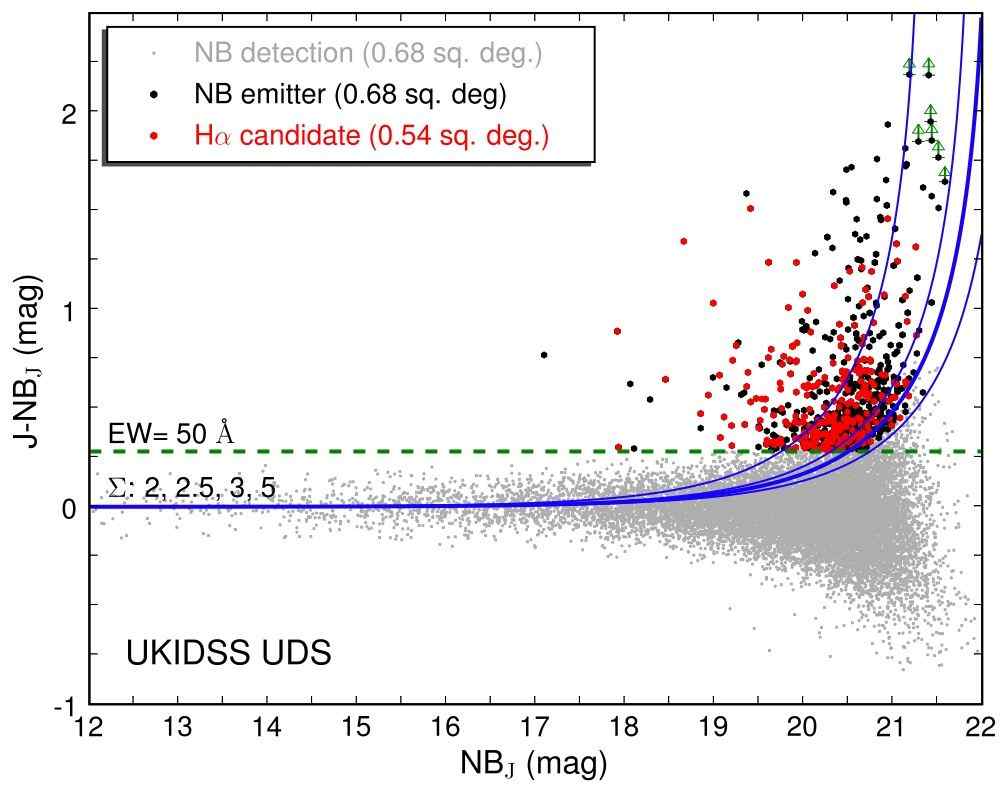}
\end{minipage}
\caption[Colour-magnitude COSMOS_UDS]{Colour-magnitude plots for COSMOS (left panel) and UDS (right panel) showing all $>3\sigma$ detections in the NB$_J$ image. The curves represent $\Sigma$ significances of 5, 3, 2.5 and 2, respectively, as defined in the text (for an average depth). The dashed line represents an equivalent width cut of 50\AA. All selected narrow-band emitters are plotted in black, while candidate H$\alpha$ emitters (selected using photometric redshifts from Mobasher et al.\ 2007 for COSMOS and Cirasuolo et al.\ 2008 for UDS) are plotted in red. Emitters with no clear $J$-band detection are assigned a 3-$\sigma$ upper limit.\label{colour-magC}}
\end{figure*}

As the NB$_J$ filter does not fall at the centre of the $J$-band, objects with redder colours will tend to have a negative $(J-$\,NB$_J)$ colour, while bluer sources will have $(J-$\,NB$_J)>0$. This will affect the selection of emission line objects, but can be corrected for by considering the broad-band colours of each source. To
do this, $(J-$\,NB$_J)$ was plotted as a function of $(z-J)$ (COSMOS) or $(J-K)$ (UDS) colour and a linear fit was derived, determining trends with a slope of $(J-$\,NB$_J)\sim 0.06 (J-K)$ for UDS and  $(J-$\,NB$_J)\sim 0.11 (z-J)$ for COSMOS. This was then used to correct the NB$_J$ magnitudes and thus the resultant $(J-$\,NB$_J)$  colours. Sources with no $z$ or $K$ data -- either because they are too bright or too faint -- were corrected assuming a median colour (based on sources at similar magnitudes).

The flux of an emission line, F$_{{\rm line}}$, and the equivalent width, EW$_{{\rm line}}$, can then be expressed as:
\begin{equation}
   {\rm F}_{{\rm line}}=\delta_{NB_J}\frac{f_{NB_J}-f_J}{1-(\Delta\lambda_{NB_J}/\Delta\lambda_{J})}
\end{equation}
and
\begin{equation}
   {\rm EW}_{{\rm line}}=\delta_{NB_J}\frac{f_{NB_J}-f_J}{f_J-f_{NB_J}(\Delta\lambda_{NB_J}/\Delta\lambda_{J})}
\end{equation}
where $\Delta\lambda_{NB_J}$ and $\Delta\lambda_{J}$ are the widths of the broad and narrow-band filters, and $f_{NB_J}$ and $f_{J}$ are the flux densities measured for each band. EW$_{{\rm line}}$ is simply the ratio of the line flux and continuum flux density.

The selection of emission-line candidates is done imposing two conditions. Firstly, the NB$_J$ sources are considered candidate emitters if they present a colour excess significance of $\Sigma>2.5$. Secondly, only NB$_J$ detections with EW$_{{\rm line}}>50$\AA \ (corresponding to $(J-$\,NB$_J)>0.3$) are selected. This is done to avoid including bright foreground objects with a large significance and a steep continuum across the $J$ band \citep[e.g.][]{2000A&A...362..509V}, and was chosen to reflect the general scatter around the zero colour a bright magnitudes. A comparison with spectroscopic data (see \S\ref{red}) indicates that these criteria maximise the completeness of the sample without introducing a significant number of false emitters.  Figure~\ref{colour-magC} shows the colour-magnitude diagrams with the selection criteria for COSMOS and UDS, respectively. Figure~\ref{excess} presents two examples of emitters.

\subsection{Complete sample}

Narrow-band detections below the 3-$\sigma$ threshold were not considered. When there was no $>3\sigma$ $J$-band detection, a 3-$\sigma$ upper limit for the $(J-$\,NB$_J)$ colour was computed, as shown in Figure~\ref{colour-magC}. The average 3-$\sigma$ line flux limit over both fields is $8\times10^{-17}$\,erg\,s$^{-1}$\,cm$^{-2}$.
The complete sample has 1517 excess sources out of all 37222 NB$_J$ detections in the entire area, with 824 being detected in COSMOS (corresponding to 1084 emitters per $\deg^2$) and 693 in UDS (1020 emitters per $\deg^2$). The  potential emitters were all visually inspected in both $J$-band (when detected) and NB$_J$, and 48 were removed from the sample as they were flagged as spurious. The majority of these (34) correspond to artifacts caused by bright stars that are on the edges of two or more frames simultaneously. The remaining 14 sources removed were low S/N detections in noisy regions of the NB$_J$ image.

\subsection{Photometric redshift analysis} \label{foto}

In order to select H$\alpha$ emitters at $z=0.84$, one needs to separate them from other line emitters at different redshifts. The detection of H$\beta$ and [O{\sc iii}]\,5007 emitters at $z\sim1.4$--1.5 is expected, together with [O{\sc ii}]\,3727 at $z=2.23$ and other emission lines. Multi-wavelength data, photometric and spectroscopic redshifts available for both COSMOS \citep{2007ApJS..172..117M,zCOSMOS} and UDS \citep{2008arXiv0804.3471C} are therefore used to distinguish between different emission lines and also to evaluate how robust the emitter selection criteria are. In order to do this, photometric redshifts for the emission line candidates were taken from the photometric redshift catalogues for COSMOS and UDS \citep{2007ApJS..172..117M,2008arXiv0804.3471C}. For UDS, the overlap with the photometric redshift catalogues reduces the area coverage to $0.54\deg^2$ due to the required overlap with Subaru optical data and further masking around bright stars: 133 sources are ``lost'' as they are in the excluded area. In addition, for 14 sources photometric redshifts are not available in the catalogues due to i/$K_s$ magnitude limits (3 in COSMOS, 11 in UDS): while these appear to be real sources (although there is a chance they are spurious), they are likely to be at redshifts higher than $z=0.84$ and so are excluded from this analysis of H$\alpha$ emitters. The photometric redshift matches result in a final sample of 1370 excess sources (821 in COSMOS, 549 in UDS) meeting all of the selection criteria across a total of $1.30\deg^2$.

%
%
\begin{figure*}
\begin{minipage}[b]{0.48\linewidth} 
\centering
\includegraphics[width=8.2cm]{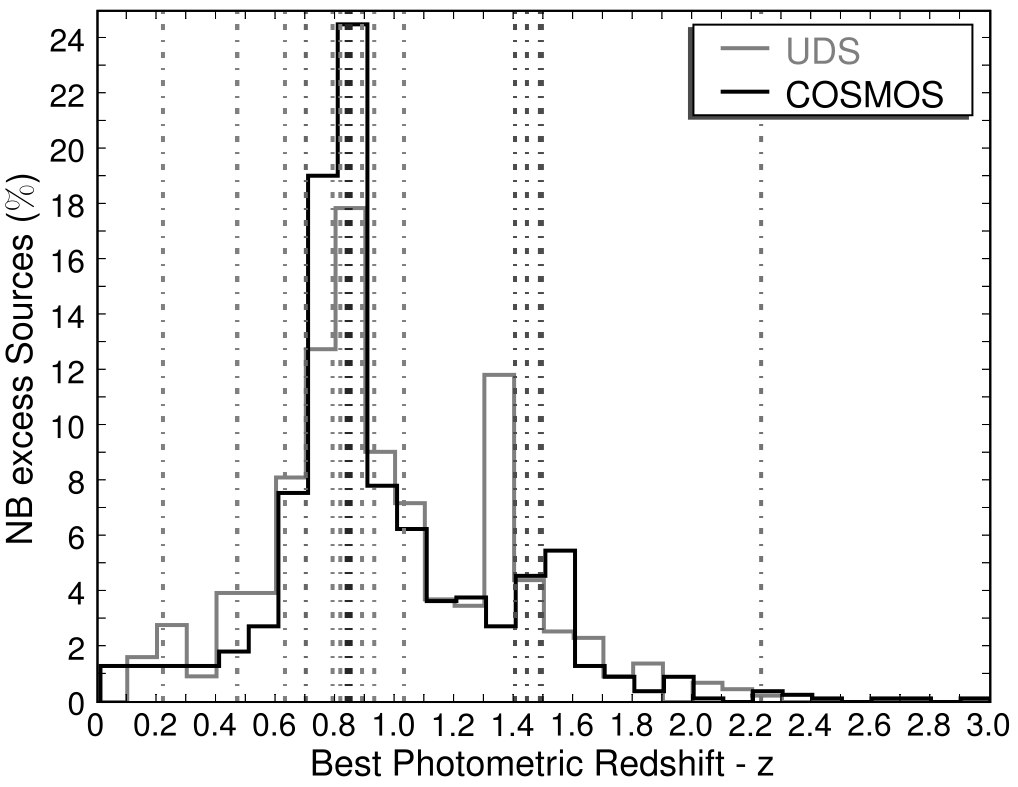}
\end{minipage}
\hspace{0.1cm} 
\begin{minipage}[b]{0.48\linewidth}
\centering
\includegraphics[width=8.2cm,height=6.3cm]{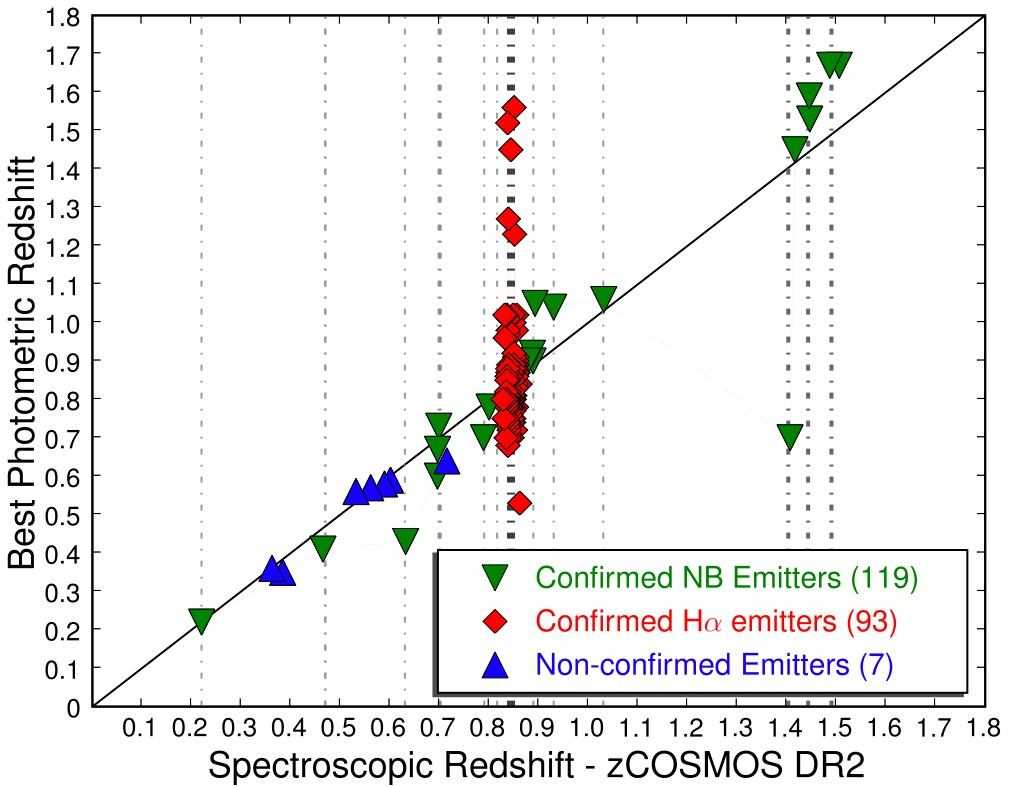}
\end{minipage}
\caption[Spectroscopic-photo]{The left panel shows the photometric redshift distribution (peak of the probability function for each source) for the selected excess sources for COSMOS and UDS. Both distributions peak at $z \sim 0.85$, as expected, thus implying that the great majority of the selected sources are H$\alpha$ emitters. The right panel shows the best photometric redshifts plotted against the spectroscopic redshifts for the emitters with accurate redshifts (reliability $> 95$\%) measured in the $z$-COSMOS DR2 dataset. Dash-dotted lines indicate the redshifts at which emission lines are matched with the NB$_J$ filter and these are the same in both figures (the width of the lines scales with the fraction of confirmed emitters; see Table~\ref{z_table} for more details). The great majority of potential emitters is confirmed to be real and only a small fraction ($\sim 6 \%$) was not confirmed. \label{spectroz}}
\end{figure*}

The photometric redshifts in COSMOS are particularly good for selecting H$\alpha$ emitters at $z=0.84$ as the available data are ideal to probe the 4000\AA\ break at that redshift. The deeper near-infrared data used in the UDS photometric redshifts  provide better accuracy for higher redshift sources, and are accurate enough to probe the proportion of $z\sim1.4$--1.5 [O{\sc iii}]\,5007 and H$\beta$ emitters. Photometric redshifts for COSMOS present $\sigma(\Delta z ) = 0.03$, where $\Delta z$ = $(z_{phot}-z_{spec})/(1+z_{spec})$. The fraction of outliers, defined as sources with $\Delta  z>3\sigma(\Delta z)$, is lower than 3\%, based on results from $z$-COSMOS (see \S\ref{red}). For UDS, the photometric redshifts have $\sigma(\Delta z) = 0.04$, with 2\% of outliers.

Figure~\ref{spectroz} shows the photometric redshift distribution for the selected narrow-band emitters in COSMOS and UDS, demonstrating very good agreement between the two fields, despite the completely different photometric analyses -- done with different codes, bands and by completely independent teams. Both samples peak at $z\sim0.85$, implying that the majority of the narrow-band excess sources are indeed H$\alpha$ emitters. In addition to this, there is another peak at $z\sim1.4$--1.5 in UDS and the same -- but at a slightly higher photo-z -- in COSMOS. This is interpreted as a significant population of [O{\sc iii}]\,5007 emitters at $z\sim1.4$ and/or H$\beta$ emitters at $z\sim1.5$. However, the photometric redshifts are not accurate enough to clearly distinguish between these two populations, and thus the slight difference between UDS and COSMOS might be just caused by the use of different bands and methods giving different performance at $z>1$.

%
%
\begin{table*}
 \centering
  \caption{Spectroscopic redshifts matches (from $z$-COSMOS DR2) for the narrow-band excess objects. The first column of the  table indicates the emission line producing the narrow-band excess, based on the accurate spectroscopic redshifts which place such lines within the narrow-band filter. The remaining columns present the mean redshift, the number of emitters and the number of those selected as H$\alpha$ using the photometric redshift selection (see \S3.4).}
  
  \begin{tabular}{@{}ccccc@{}}
  \hline
   Emission Line & $\lambda$ (\AA) & $<z>$ & Number & Selected as H$\alpha$ \\
 \hline
H$\alpha$ & 6563 & 0.845 & 93  & 88 \\
 \hline
H$\beta$ & 4861 & 1.49 & 2 & 0 \\
$[$O{\sc iii}$]$ & 5003 & 1.42 & 4 & 1 \\
He{\sc i} & 5876 & 1.04 & 1 & 0  \\
$[$O{\sc i}$]$ & 6363 & 0.89 & 4  & 0 \\
$[$S{\sc ii}$]$ & 6717 & 0.79 & 2  & 1 \\
$[$Ar{\sc iii}$]$ & 7135 & 0.70 & 3 & 0 \\
$[$O{\sc ii}$]$ & 7325 & 0.63 & 1  & 0 \\
He{\sc ii} & 8237 & 0.47 & 1 & 0 \\
C{\sc i} & 9830 & 0.22 & 1 & 0 \\
Unidentified & &  & 7 & 0 \\
 \hline
\end{tabular}
\label{z_table}
\end{table*}

\subsection{Selecting H$\alpha$ emitters at ${z=0.84}$} \label{selecting}

The sample of candidate H$\alpha$ emitters at $z=0.84$ is derived by considering not only the best-fit photometric-redshift for each source, but also more extensive information contained within the photometric redshift probability distribution. In particular, the H$\alpha$ candidates are defined to be those sources with $z_{min}<0.845<z_{max}$ (where $z_{min}$ and $z_{max}$ are the 1-$\sigma$  redshift limits of the principle peak in the photometric redshift probability distribution). For COSMOS, the photometric redshift selection produces a sample of 477 potential H$\alpha$ emitters, while for UDS the same procedure yields 270 sources (see \S\ref{foto} for details).

%
%
\begin{figure*}
\begin{minipage}[b]{0.48\linewidth} 
\centering
\includegraphics[width=8.2cm]{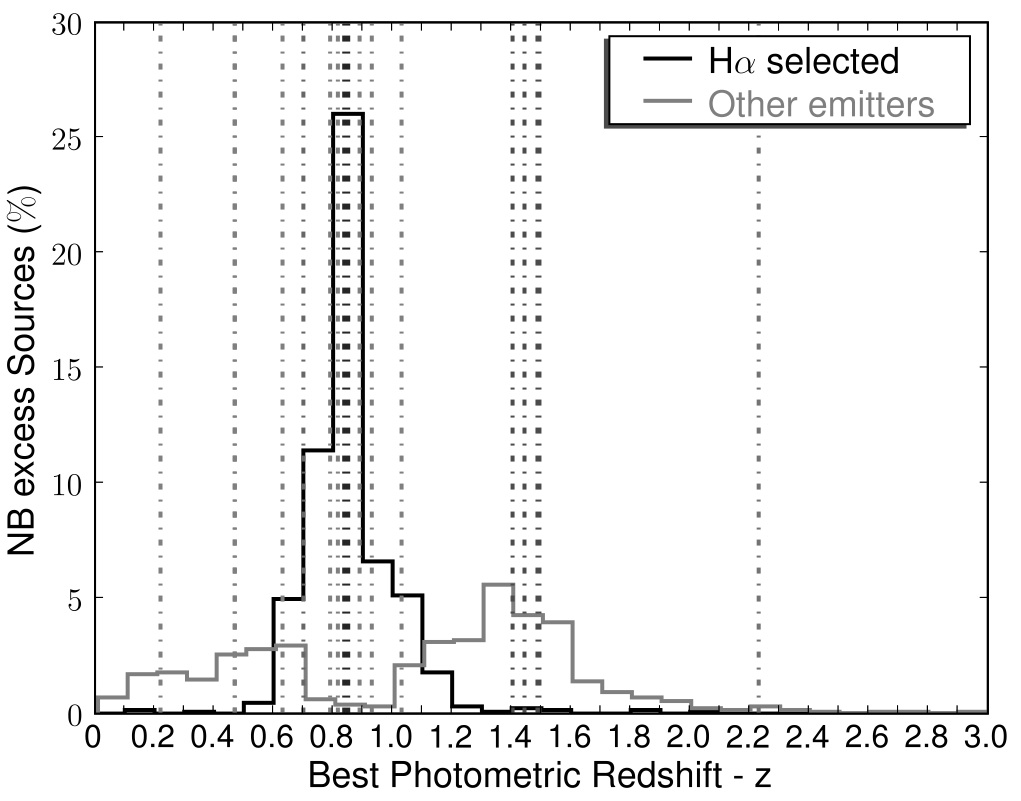}
\end{minipage}
\hspace{0.1cm} 
\begin{minipage}[b]{0.48\linewidth}
\centering
\includegraphics[width=8.2cm,height=6.3cm]{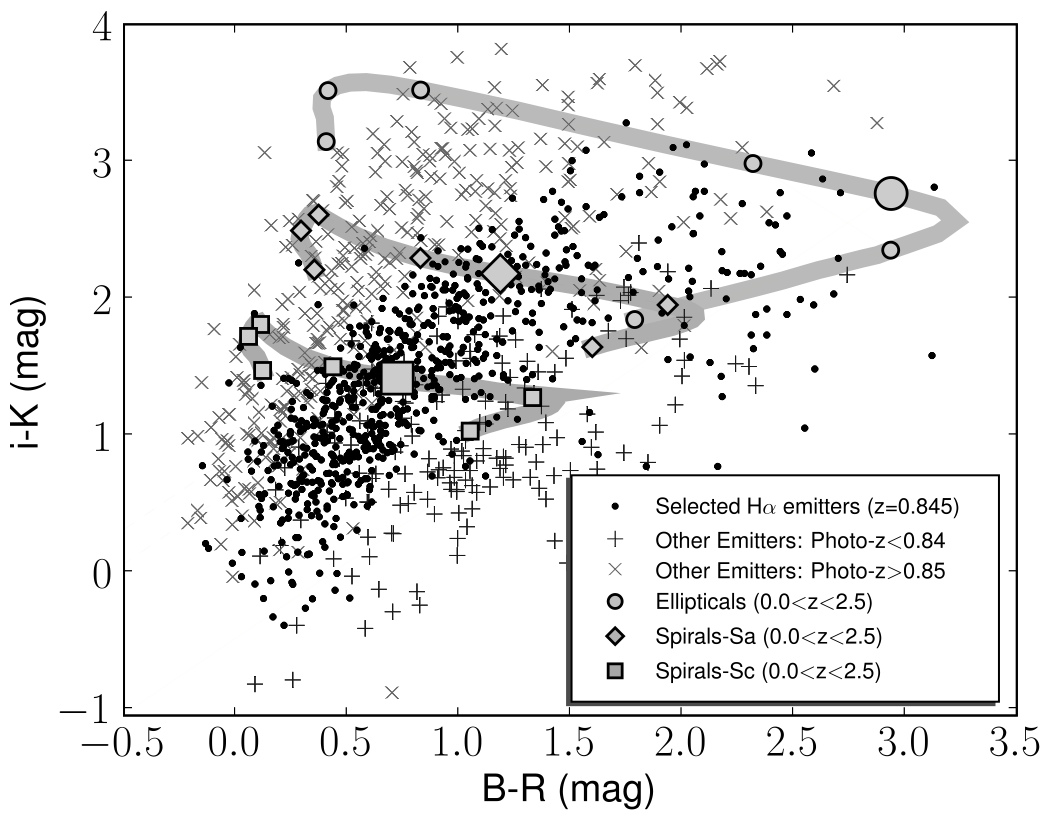}
\end{minipage}
\caption[Spectroscopic-photo]{The left panel shows the photometric redshift distribution (peak of the probability function for each source) for the selected H$\alpha$ emitters (see Section \ref{selecting}) and the remaining emitters, with the spectroscopically confirmed lines plotted as in the left panel of Figure \ref{spectroz}. The right panel presents a BRIK colour-colour diagram, clearly distinguishing the high-redshift emitters from the low-redshift emitters, and also the H$\alpha$ emitters from these two populations. Overlaid on the plot are evolutionary tracks for different galaxy
models, drawn from \cite{1997A&AS..122..399P}; small symbols on the tracks mark steps of 0.5 in redshift from 0 to 2.5, whilst the larger symbol
indicates z=0.84. The observed colours for the H-alpha emitters are consistent with them being mostly star-forming galaxies at z=0.84, whilst the colours of the other emitters also lie in the expected regions. \label{spectroz2}}
\end{figure*}

While it is true that this selection can potentially introduce some
biases, a careful comparison of the selected H$\alpha$ emitters and the
remaining emitters shows no evidence for such an effect. The emission line
equivalent width distributions of the selected H$\alpha$ and non-H$\alpha$
emitters are statistically indistinguishable. The sample of non-selected
emitters does contain objects which have slightly fainter magnitudes (and
correspondingly fainter emission lines) than the selected H$\alpha$
emitters, but the colours of these fainter galaxies are clearly consistent
with them simply being at higher redshift (e.g.\ [OIII], H$\beta$ or [OII]
emitters): they are typically detected at a $\sim$20-30 $\sigma$ (or higher) level in
the optical images, and thus it should not be any inaccuracy in the
photometric redshifts which affects their selection. One can also see that the photometric redshift distribution in Figure \ref{spectroz2} (left panel) is clearly different between the selected H$\alpha$ emitters and the remaining emitters, revealing both a population of lower and higher redshift. The right panel of Figure \ref{spectroz2} also reveals that the H$\alpha$ selected emitters present colours which clearly distinguish them from the low-z and higher-z samples of emitters -- they also occupy a region in the colour-colour diagram where one expects to find star-forming galaxies at z=0.84. Finally, as examined below, spectroscopic redshifts for $\sim$ 25\% of the sample in COSMOS show that only a negligible fraction of the H$\alpha$ emitters are being missed by the photometric redshift selection.

\subsection{Spectroscopic redshifts and selection robustness}
\label{red}

Both COSMOS and UDS have large spectroscopic surveys underway, but only a limited fraction of those spectroscopic redshifts is currently available. Nevertheless, for the $z$-COSMOS survey \citep{zCOSMOS} Data Release 2 (DR2), the match with the sources detected on the NB$_J$ frames yields 4600 sources (using a $1.0''$ match criteria), of which 138 (each matching only one $z$-COSMOS source) have been selected as excess sources  and hence potential emitters. Twelve of these have unreliable or non-existent redshift determinations. Of the remaining 126, the vast majority (119) have redshifts which place an emission line within the NB$_J$ filter (see Table \ref{z_table} for the detailed list of emitters), confirming the narrow-band excess as an emission line. For the other 7 sources, it was not possible to identify any emission line falling into the NB$_J$ filter. Spectra for these sources were analysed in order to look for potential errors in the redshift determination. In fact, while redshifts for 5 of them are very robust, one source, identified as $z\sim0.4$, seems better fitted by being at $z=0.85$ (with [O{\sc ii}]\,3727 being detected, while the fit at lower redshift assumes an emission line where fringing starts to become an issue in the spectrum). The remaining source contains one emission line (identified as [O{\sc ii}]\,3727) but with a low signal to noise (S/N\,$<2.0$). These 7 sources all have a colour excess significance $\Sigma=2.5$--3.0, and at least the 5 robust cases represent the galaxies expected to be randomly scattered into the candidate emitter list at these low $\Sigma$ values. All of the candidate emitters with $\Sigma>3.0$ were confirmed to be real. It is also worth noticing that $z$-COSMOS DR2 is highly biased towards $z<1$ sources, which means that its completeness for potential interlopers (which are more likely to be in this redshift range) is very high, while at the same time it misses most of the emitters at higher redshift. Therefore, the contamination rate within the sample of emitters (i.e., the fraction of non-emitters) is likely to be lower than $\sim6$\%, even down to $\Sigma=2.5$.

Tests regarding the selection criteria used were also done by selecting samples with different colour significances and different equivalent width cuts, and comparing the matches for each selection criteria with $z$-COSMOS DR2. As expected, the number of potential interlopers declines rapidly with an increasing colour significance threshold, but so does the sample size. The sample with the largest number of real emitters whilst returning a low ratio between potential interlopers and real emitters is obtained using $\Sigma\sim2.5$ and an equivalent width cut of $\sim50$\AA, therefore indicating that the selection criteria used are producing reliable results.

Table \ref{z_table} compares the spectroscopic selection of line emitters with the selection from the photometric redshift catalogue. It can be seen that 88 of the 93 H$\alpha$ emitters are correctly identified  ($\sim95$\% completeness). On the other hand, 2 non-H$\alpha$ emitters (with fluxes which do not deviate significantly from the median) were found in the photometric redshift selected sample (implying a $\sim98$\% reliability), with one of those being a [S{\sc ii}]\,6731 emitter. Because this line is very close to H$\alpha$ it is very difficult to completely distinguish these emitters with photometric redshifts alone. However, as seen by the sample which has been presented, the total contamination by these will be very small ($<2$\%). The spectroscopic data were used to improve the H$\alpha$ catalogue, removing the 2 non-H$\alpha$ emitters and including those 5 which had not been selected by the photometric redshifts. These few sources present mean and median fluxes slightly higher than the rest of the sample, suggesting that some bright H$\alpha$ emitters could be lost by a photometric redshift selection, due to the emission line contribution to the broad-band fluxes. However, as the brighter H$\alpha$ emitters are likely to contain strong emission lines in the visible ([O{\sc ii}]\,3727, H$\beta$ and [O{\sc iii}]\,5007), these targets should have a very high completeness in $z$-COSMOS, so this can be interpreted as an upper limit. As these emitters are introduced into the sample, no further specific correction was applied.

While 93 H$\alpha$ emitters are confirmed using $z$-COSMOS, there are 72 $z$-COSMOS galaxies with a spectroscopic redshift which should place the H$\alpha$ line in the NB$_J$ filter. The failure to select these as narrow-band excess sources means that they have a weak or absent H$\alpha$ emission line. These present a mean colour $(J-$\,NB$_J)=0.1$ corresponding to a mean measured H$\alpha$ flux lower than the survey limit ($3\times 10^{-17}$\,erg\,s$^{-1}$\,cm$^{-2}$ and a mean EW of 20\AA). The brightest source has $J=19$ mag. These are likely to be very faint emitters for which a completeness correction is applied later (see \S\ref{compl}). The [O{\sc ii}]\,3727 line fluxes from $z$-COSMOS galaxies confirms this: where they are detected they yield similar line fluxes to the ones that would be estimated for H$\alpha$. On the other hand, some of these sources have a negative $(J-$\,NB$_J)$ colour. By applying similar selection criteria used for emitters, it seems that at least 2 H$\alpha$ absorbers would be selected.

\subsection{Narrowband-$K$ and $H$ matches}

By design, the custom-built narrow-band filters make it possible to look for line detections in multiple bands to refine redshift estimates. Currently, data are available in
H$_{2}$S$_{1}$ for COSMOS (Geach et al.\ 2008) and NB$_H$ and H$_{2}$S$_{1}$ for UDS (Sobral et al., Geach et al., in prep.)

For COSMOS, it was possible to match 3 sources between excess sources catalogs ($J$ and $K$), confirming a $z=2.23$ redshift for those emitters. This shows that the line being
detected in NB$_J$ is [O{\sc ii}]\,3727.

For UDS, 18 sources were matched between the NB$_H$ and NB$_J$. These have photometric redshifts consistent with either being H$\alpha$ emitters at $z\sim 1.5$ or H$\alpha$ emitters at $z=2.23$, thus meaning that the line detected in the $J$ band is H$\beta$ or [O{\sc ii}]\,3727 (respectively)  for these emitters. At least 2 of these are also selected as emitters in the NB$_K$ filter, which confirms $z=2.23$ for them. In addition, there are another 6 matches between the $K$ and the $J$ narrow-band filters, indicating that those sources are at  $z=2.23$ as well. Four of these 24 matched sources were found in the photometric redshift-derived H$\alpha$ catalog and were thus removed as the evidence points towards these being at higher redshift.

\section{RESULTS}

A total of 743 candidate $z = 0.84$ H$\alpha$ emitters are detected over $1.30\deg^2$ down to an average observed line flux limit of $8\times10^{-17}$\,erg\,s$^{-1}$\,cm$^{-2}$. This sample will now be used to evaluate the H$\alpha$ luminosity function and estimate the star formation rate density at $z=0.84$. The morphological mix of these H$\alpha$ emitters in the COSMOS field will then be investigated, together with their evolution as a function of luminosity and redshift, and their contribution to the total luminosity function and star formation rate density.

\subsection{H$\alpha$ luminosity function at $z=0.84$} \label{lfalpha}

In order to calculate the luminosity function of H$\alpha$ emitters, line fluxes are converted to luminosities by applying:

\begin{equation}
  L_{{\rm H_{\alpha}}}=4\pi {\rm D_L^2}{\rm F}_{{\rm H_{\alpha}}}
\end{equation}
where ${\rm D_L}$ is the luminosity distance, 5367\,Mpc at $z=0.84$.

The estimate of the source density in a luminosity bin of width $\Delta(\log L)$ centered on $\log L_c$ is given by the sum of the inverse volumes of all the sources in that bin. Therefore, the value of the source density in that bin is

\begin{equation}
 \phi(\log(L_c))=\frac{1}{\Delta(\log L)} \sum_{|\log \frac{L_i}{L_c}| < \frac{\Delta(\log L)}{2}} \frac{1}{\Delta(V_{\rm filter})}.
\end{equation}
Here, $i$ refers to sources and $c$ to the center of each bin. The volume probed is calculated taking into account the survey area and the narrow-band filter width (initially assumed to be a top-hat function across 1.2037--$1.2185\umu$m; see \S\ref{profil_filt} for a refined approach), resulting in a co-moving volume of $7.4 \times 10^4$\,Mpc$^{3}$ (UDS) and $1.04 \times 10^5$\,Mpc$^{3}$ (COSMOS). As detailed before, this takes into account the removed area due to the presence of bright stars and consequent artifacts, along with noisy areas.

The luminosity functions presented here are fitted with a Schechter function defined by the three parameters: $\alpha$, $\phi ^*$ and $L^*$: 

\begin{equation}
  \phi(L)dL=\phi^*(L/L^*)^\alpha \exp(-L/L^*)d(L/L^*).
\end{equation}

\subsubsection{[NII] flux contamination correction}
\label{nii}

When computing line fluxes and equivalent widths for the H$\alpha$ line, one must note that the adjacent [N{\sc ii}] lines at 6548 and 6583\AA\ will also contribute to both quantities, increasing them both. In order to account for this, the relation between the flux ratio F$_{{[\rm N\sc II]}}/$F$_{{\rm H}\alpha}$ and the total measured equivalent width EW(H$\alpha$+[N{\sc ii}]) derived by Villar et al.\ (2008) is used. This shows that the fractional [N{\sc ii}] flux contribution decreases with increasing EW. The relation between $y=\log($\,F$_{{\rm {[N}{\sc II}]}}/$\,F$_{{\rm H}\alpha})$ and $x=\log($\,EW(H$\alpha+[$N{\sc ii}$]))$ can be approximated by $y=-5.78+7.63x -3.37x^2+0.42x^3$ for EW\,$>50$\AA. This results in a correction lower than the conservative 33\% used by some authors (e.g.\ Geach et al.\ 2008), with the median being 25\% for this sample with EW\,$>50$\AA.

\subsubsection{Completeness correction}
\label{compl}

Fainter sources and those with weak emission lines might be missed and thus not included in the sample; this will result in the underestimation of the number of emitters, especially at lower luminosities. Furthermore, the completeness rate is highly affected by the selection criteria ($\Sigma>2.5$, EW\,$>50$\AA, $\sigma>3$). To address this problem, a series of simulations were conducted. First, the recovery rate has been studied as a function of magnitude for objects in the same frame, using a Monte Carlo method. For this, 10 different galaxies (both real -- taken from the narrow-band images in UDS and COSMOS -- and simulated) were used. These were introduced (20 for each run, in any given image) with different input magnitudes into the science frames and then detected using the same extraction parameters as the main catalogues. The recovery rate and the recovered magnitude were then studied. The latter follows the input magnitude reasonably well down to the 3-$\sigma$ limit. The recovery rate, while varying slightly with the type of galaxy used (the artificial and point-like objects, for example, showed a much higher recovery rate at fainter levels than real galaxies), falls off sharply fainter than NB$_J\sim21$. Similar simulations were also done for the $J$ band images.

%
%
\begin{figure}
\centering
\includegraphics[width=8.2cm]{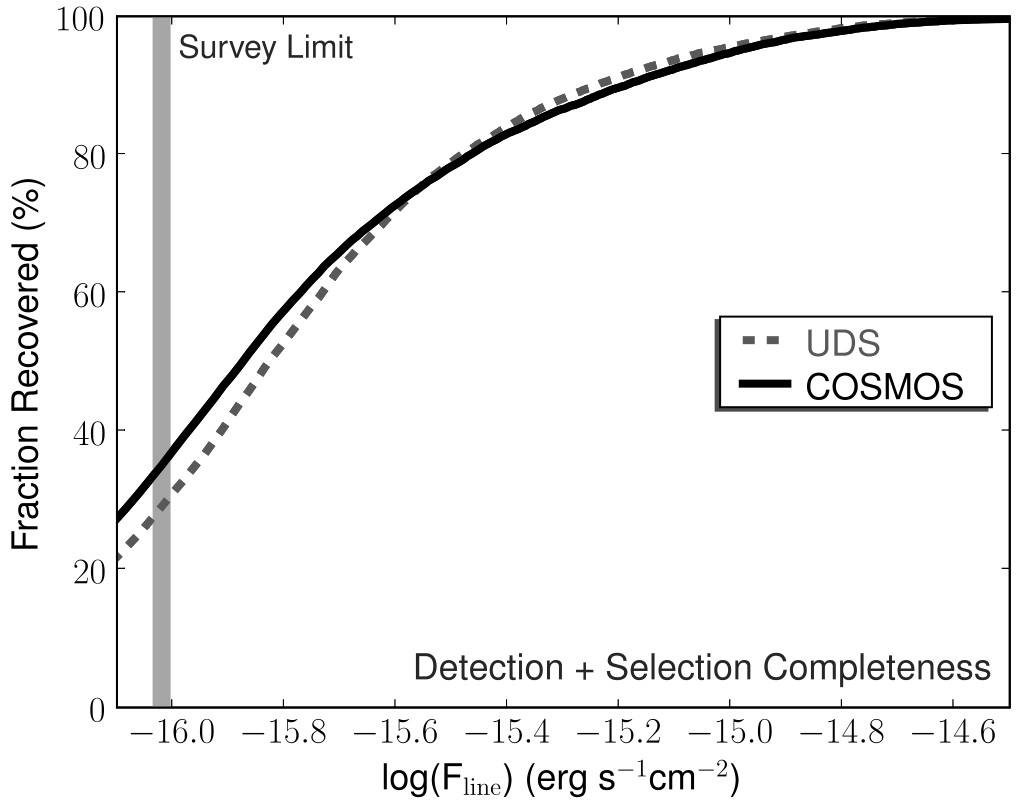}
\caption[Detection, line recovery completeness and magnitude tests]{The total completeness function used for COSMOS and UDS resulting from several simulations to address the effects of selection and detection. This confirms a $\sim 30$\% completeness at a flux limit of $\sim 10^{-16}$\,erg\,s$^{-1}$\,cm$^{-2}$.\label{completeness}}
\end{figure}

This information alone cannot be used to correct the luminosity function, as faint narrow-band sources don't necessarily have faint H$\alpha$ lines and vice-versa. To address this, a second series of simulations were performed to investigate the selection completeness of faint emission lines within detected galaxies. For this, narrow-band detections not classified as emitters were considered not to have any emission line. For each of those sources ($\sim21000$ for COSMOS and $\sim15000$ for UDS) a line flux (in steps of $5\times10^{-18}$\,erg\,s$^{-1}$\,cm$^{-2}$ up to $5\times10^{-15}$\,erg\,s$^{-1}$\,cm$^{-2}$) was added to both NB$_J$ and $J$-band magnitudes, and the selection criteria were applied to the revised magnitudes for each line flux to study the recovery rate. This outputs a colour selection completeness for each line flux. Finally, a combined completeness correction was done combining both simulations described. The combined completeness function, which is used to correct the luminosity function on a bin to bin basis, is shown in Figure~\ref{completeness}. Slightly higher completeness rates at the same flux level have been claimed in \cite{2008ApJ...677..169V}, for shallower data, but it should be pointed out that these authors use a combination of selection criteria which is significantly different from those used in this study. On the other hand, they only consider the colour-selection completeness, neglecting the detection completeness, and on the other hand, they use sources detected down to just $0.8\sigma$ in NB$_J$ (instead of the 3-$\sigma$ limit used in this work), and with a lower EW ($\sim15$\AA) cut; therefore, the results presented here are likely to be more reliable than those by \cite{2008ApJ...677..169V}.

%
%
\begin{table*}
 \centering
  \caption{The H$\alpha$ luminosity function. Superscript numbers in the column titles indicate the corrections made in each stated value: 1 -- extinction correction ($A_{H\alpha}=1$\,mag) and [N{\sc ii}] correction as a function of total measured EW; 2 -- completeness correction; 3 -- filter profiles biases correction.}
  \begin{tabular}{@{}ccccccc@{}}
  \hline
   log(L$_{H\alpha})$ & \# Sources$^{1}$  & log($\Phi$)$^{1}$ & log($\Phi$)$^{1,2}$ & log($\Phi$) final$^{1,2,3}$ & $\Delta$log($\Phi$)$^{1,2,3}$ \\
   (erg s$^{-1}$) &  & (Mpc$^{-3}$) & (Mpc$^{-3}$) & (Mpc$^{-3}$) & (Mpc$^{-3}$) \\
 \hline
      \noalign{\smallskip}
41.7 (41.65--41.75) &  101 & $-$2.25 & $-$1.77 & $-$1.77 & 0.14\\
41.8 (41.75--41.85) & 133 & $-$2.13 & $-$1.78 & $-$1.74 & 0.10\\
41.9 (41.85--41.95) & 132 & $-$2.13 & $-$1.88 & $-$1.83 & 0.10\\
42.0 (41.95--42.05) & 110 & $-$2.20 & $-$2.04 & $-$1.96 & 0.10\\
42.1 (42.05--42.15) & 79 & $-$2.35 & $-$2.23 & $-$2.12 & 0.11\\
42.2 (42.15--42.25) & 69 & $-$2.41 & $-$2.32 & $-$2.24 & 0.11\\
42.3 (42.25--42.35) & 39 & $-$2.66 & $-$2.59 & $-$2.48 & 0.14\\
42.4 (42.35--42.45) & 24 & $-$2.87 & $-$2.81 & $-$2.68 & 0.18\\
42.5 (42.45--42.55) & 14 & $-$3.10  & $-$3.06  & $-$2.92 & 0.22\\
42.6 (42.55--42.65) & 9 & $-$3.30 & $-$3.26 & $-$3.11 & 0.27\\
42.7 (42.65--42.75) & 8 & $-$3.50 & $-$3.45 & $-$3.35 & 0.28\\
42.9 (42.75--43.05) &  6 & $-$4.05 & $-$4.04  & $-$4.02 & 0.37\\
 \hline
\end{tabular}
\label{mass_table}
\end{table*}

\subsubsection{Extinction Correction}

It is well-known that the H$\alpha$ emission line is not immune to dust extinction, although it is considerably less affected than Ly$\alpha$ or the UV continuum. Measuring the extinction for each source can in principle be done by several methods, ranging from spectroscopic analysis of Balmer decrements to a comparison between H$\alpha$ and far-infrared determined SFRs. For now, however, a conservative A$_{H\alpha}=1$ mag is used, which is the same correction adopted in most of the similar studies done before \citep[e.g.][]{2003ApJ...586L.115F, 2007ApJ...657..738L,2008arXiv0805.2861G}. This same correction is also applied to all other datasets being compared here, when possible, in order to give consistency between all studies and search for evolution. \S\ref{aign} shows that this correction should not be too far from the actual mean extinction for the entire sample.

\subsubsection{Filter profile corrections} \label{profil_filt}

The narrow-band filter transmission function is not a perfect top-hat (as assumed earlier), and thus the real volume probed varies as a function of intrinsic luminosity: luminous H$\alpha$ emitters will be detected over a larger volume than the fainter ones because they can be detected in the wings of the filters (although they will be detected as fainter sources in these cases). Low luminosity sources will only be detected in the central regions of the filter and thus the effective volume will be smaller.

In order to correct for this when deriving the H$\alpha$ luminosity function, a further set of simulations was run. Firstly, the luminosity function was computed with the corrections described above and the best fit was assigned. This was then assumed to be the true luminosity function, allowing the generation of a set of $\sim 10^5$ H$\alpha$ emitters with a flux distribution given by the measured luminosity function, but spread evenly over the redshift range $z=0.81$--0.87 (assuming no cosmic structure variation or evolution of the luminosity function over this narrow redshift range). The top-hat filter model was then confirmed to recover the input luminosity function perfectly. Next, the true filter profiles were used to study the recovered luminosity function. These simulations showed that the number of brighter sources is underestimated relative to the fainter sources. A mean correction factor between the input luminosity function and the one recovered (as a function of luminosity) was then used to correct each bin. The simulation was run again with the new luminosity function, confirming that the recovered luminosity function is very similar to the input luminosity function.

The filter profiles were also checked against the spectroscopic redshift distribution from $z$-COSMOS DR2. By distributing the artificial H$\alpha$ emitters to reproduce the luminosity function, one can predict the redshift distribution of observed sources for each filter. Those matched well the redshift distribution of the $z$-COSMOS emission line sample, although this sample only contains $\sim100$ emitters at the moment.

\subsubsection{Fully corrected luminosity function} \label{lf}

\begin{figure*}
\begin{minipage}[b]{0.48\linewidth} 
\centering
\includegraphics[width=8.2cm]{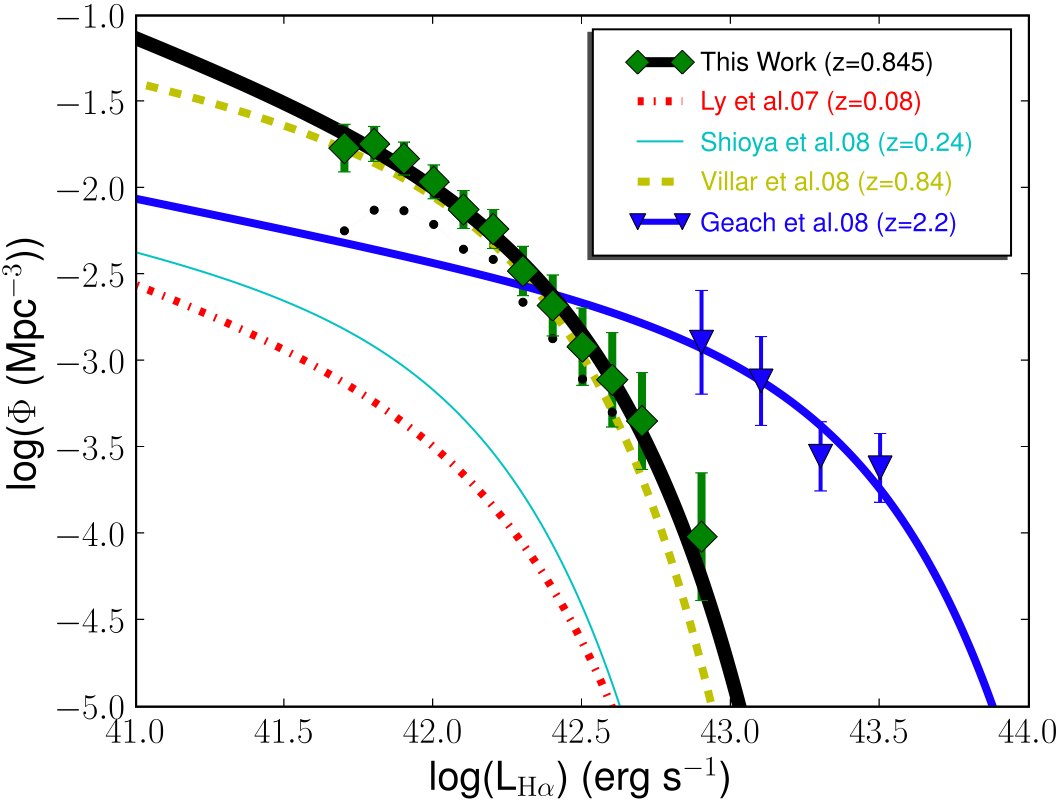}
\end{minipage}
\hspace{0.1cm} 
\begin{minipage}[b]{0.48\linewidth}
\centering
\includegraphics[width=8.2cm,height=6.3cm]{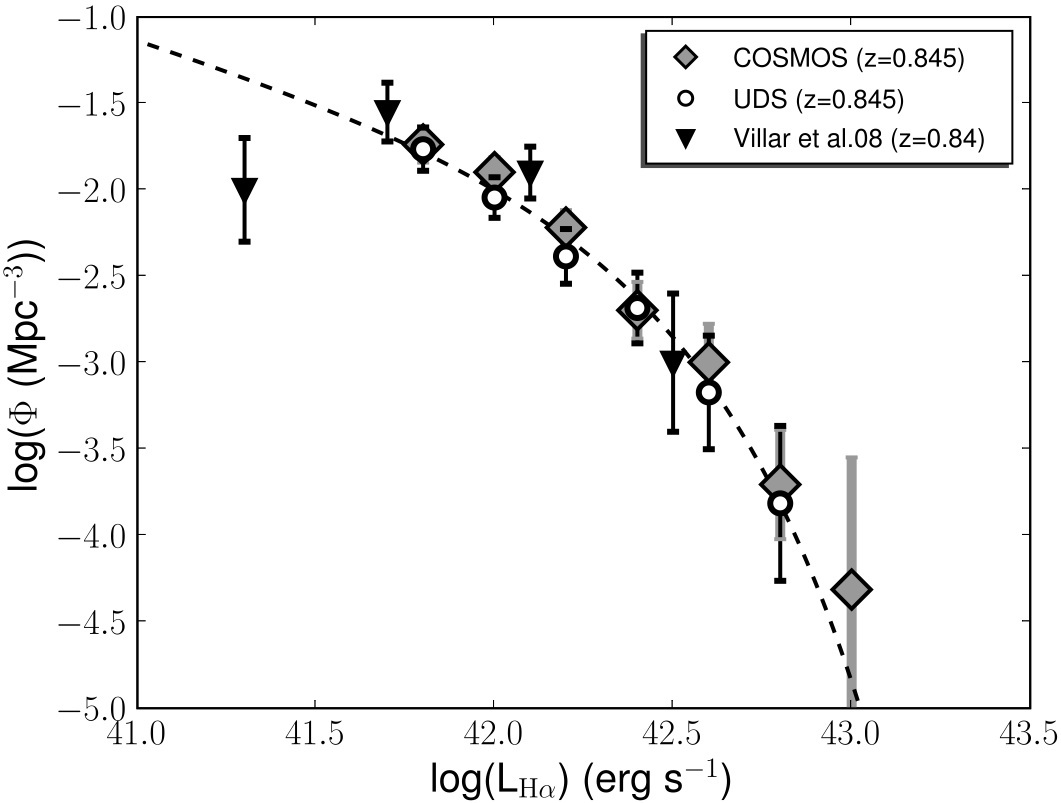}
\end{minipage}
\caption[LF]{The left panel shows the H$\alpha$ luminosity function (corrected for [N{\sc ii}] contamination, completeness, extinction and filter profile biases) with the best fit Schechter function given by $\phi^*=10^{-1.92\pm0.10}$\,Mpc$^{-3}$, $\alpha=-1.65\pm0.15$ and $L^*=10^{42.26\pm0.05}$\,erg\,s$^{-1}$. Small black dots present the derived luminosity function with corrections for [N{\sc ii}] and extinction ($A_{H\alpha}=1$ mag) only. Other luminosity functions (corrected for $A_{H\alpha}=1$, except for $z=0.08$ and $z=0.24$ where the extinction corrections from the authors were used) from narrow-band surveys of H$\alpha$ emitters at different redshifts are presented for comparison, showing a clear evolution with cosmic time up to at least $z\sim1$. The derived H$\alpha$ luminosity function for each separate field (COSMOS and UKIDSS UDS) is presented in the right-hand panel, with a comparison with data points from Villar et al. (2008) -- assuming the same dust extinction as applied in the present study. There seems to be a slightly higher density of sources in COSMOS, but both luminosity functions agree well with the combined fit (dashed line). \label{corr_lum_f}}
\end{figure*}

The final $z=0.84$ H$\alpha$ luminosity function is presented in Figure~\ref{corr_lum_f} and in Table~\ref{mass_table}. The raw luminosity function, without correcting for incompleteness and filter profile biases (correcting only for [N{\sc ii}] contamination and $A_{H\alpha}=1$ mag of extinction), is also shown. The errors are Poissonian in each bin, combined with an uncertainty on the correction factor assumed to be 10\% of the applied correction.
Luminosity functions are also computed separately for the two observed
fields (Figure~\ref{corr_lum_f}) with these being compared with the
combined luminosity function. Both luminosity functions are generally
consistent within the errors, although there seems to be a slightly higher
density at $z=0.84$ in COSMOS. The on-sky distribution of the H$\alpha$
emitters in the COSMOS and UDS fields is far from homogeneous: they are
highly clustered and several high-density regions can be found (a detailed
clustering analysis is underway and will be published in Sobral et al.\ in
preparation). Assuming a standard angular correlation function for the
H$\alpha$ emitters, parameterised as $w(\theta) = A \theta^{-0.8}$ (with $\theta$ in arcsec), then
for a value of $A \approx 1$ (typical for galaxies like
these), and following \cite{1975ApJ...196..647P}, the cosmic
variance in the number of emitters expected in a 1.3 deg$^2$ sky area due
to clustering is estimated to be roughly double the pure Poissonian
errors. This leads to (1$\sigma$)uncertainties of $\sim12$\% on the total
number of detected H$\alpha$ emitters; the corresponding uncertainties for
the individual fields are $\sim$15\% for UDS and $\sim$13\% for COSMOS.
These are fully consistent with the actual differences in number densities
found. Even for an extremely clustered population, with $A \sim 10-15$, the source count variations would not be larger than $\sim
20$\% on fields of this size. Given this, and the consistency between the
COSMOS and UDS fields (and also the results of Villar et al 2008), it is
safe to say that the results derived are largely robust against cosmic
variance.

%
%

%
%
\begin{figure*}
\centering
\includegraphics[width=17.4cm]{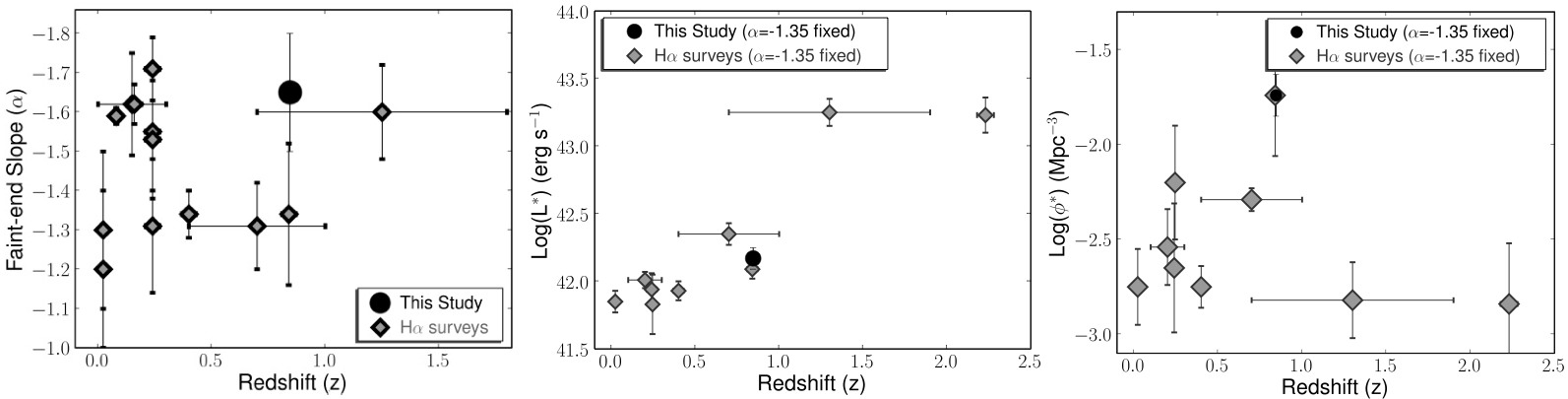}
\caption[Parameter evolution]{The evolution of the best-fit Schechter function parameters of the H$\alpha$ luminosity function for H$\alpha$ surveys. The left panel presents the best-fit values for the faint end slope, $\alpha$, which appear to fall into two distinct groups with $\alpha \sim -1.6$ and $\alpha\sim -1.3$, with little evidence for any redshift evolution. The middle and right-hand  panels show the evolution of $L^*$ and $\phi^*$ as calculated by fixing $\alpha=-1.35$. $L^*$ appears to evolve strongly with redshift out to $z>2$, whereas $\phi^*$ peaks at $z\sim1$ and then declines. Fixing the faint-end slope at $\alpha=-1.65$ yields a similar result. \label{param_evo}}
\end{figure*}

Also plotted in Figure~\ref{corr_lum_f} are other published H$\alpha$ luminosity functions at different redshift; these demonstrate a significant evolution with redshift, linked to an increase of the number density of both faint and bright emitters at least up to $z\sim1$. To quantify this evolution a Schechter function is fitted to the combined $z=0.84$ luminosity function, which yields: $\phi^*=10^{-1.92 \pm 0.10}$\,Mpc$^{-3}$, $\alpha=-1.65\pm0.15$ and $L^*=10^{42.26 \pm 0.05}$\,erg\,s$^{-1}$. These parameters, and a comparison with the best fits for COSMOS and UDS separately, can be found in Table~5. The best fit parameters indicate a steeper faint end slope at $z=0.84$ than the canonical $\alpha=-1.35$ usually assumed; similarly steep faint-end slopes for the luminosity function of star-forming galaxies have recently been found from UV studies \citep[e.g.][]{2008ApJ...686..230B}. These authors have argued for an increase in the faint-end slope of star-forming galaxies with redshift. However, evidence for a steep faint-end slope in the H$\alpha$ luminosity function ($\alpha\sim-1.6$) has been found even at $z=0.08$ by Ly et al.\ (2007) as well as in other H$\alpha$ studies at higher redshifts, $z\sim0.2$--1.3 \citep[e.g.][]{2000AJ....120.2843H,2003ApJ...586L.115F,2008AJ....135.1412D}. Figure~\ref{param_evo} shows the best-fit $\alpha$ for H$\alpha$ luminosity functions derived at $z=0.02$--1.3. This reveals no clear evidence of evolution of the faint-end slope with cosmic time; indeed, rather than having a random distribution, the values of $\alpha$ seem to cluster into two groups with $\alpha \sim-1.6$ or $\alpha \sim-1.3$. The origin of these discrepant results is unclear, but cosmic variance may play an important role.

Figure~\ref{param_evo} also shows the variation of $L^*$ and $\phi^*$ of the H$\alpha$ luminosity function out to $z\sim2$. To derive these values, only LFs of H$\alpha$ surveys which were fitted with the ``canonical" $\alpha$=$-$1.35 or that had their data points published (those were then refitted by fixing $\alpha$=$-$1.35) were used, in order to reduce degeneracies and allow a direct comparison between results. 
It should be noted that some degeneracy will still remain between L$^*$ and $\phi^*$ values. Figure~\ref{param_evo} shows a strong evolution in $L^*$, increasing by at least an order of magnitude from the local Universe (Gallego et al.\ 1995) to $z=2.23$ (Geach et al.\ 2008). The evolution of $\phi^*$ is somewhat different: while it appears to increase from $z=0$ up to $z=0.84$ (this work and Villar et al.\ 2008) by one order of magnitude, it would then need to fall at $z>1$ to be consistent with the higher redshift data of Yan et al.\ (1999) and Geach et al.\ (2008). Note that the increased value of $\phi^*$ at $z=0.84$ does not arise just due to the degeneracy between $\phi^*$ and $L^*$: a value of $\phi^*=10^{-2.7}$\,Mpc$^{-3}$ for the current data-set can be strongly rejected, having a probability $<10^{-6}$. The evolution of $L^*$ and $\phi^*$ is therefore revealing important details of the evolution of the H$\alpha$ luminosity function: from $z\sim0$ to $z\sim1$ it seems to be driven by an increase in the number density of both bright and faint emitters, with an increasing population of bright emitters at higher redshift then being responsible for a stronger evolution in $L^*$. These results are also consistent with studies done using 24 $\mu$m data \citep[e.g.][]{2007ApJ...660...97C}. The evolutionary trends do not change if $\alpha$ is fixed at a higher value ($\alpha=-1.65$ for example).

\subsection{The star formation rate density at $z=0.84$}

\subsubsection{AGN contamination} \label{aign}

The H$\alpha$ luminosity function previously derived used all the H$\alpha$ emitters from the survey, and while most of such sources are likely to be star-forming galaxies, some of these can also be AGN. Spectra from $z$-COSMOS DR2 were used to explored this. A visual inspection of the 93 available spectra was done to confirm additional emission lines ([O{\sc ii}]\,3727, [O{\sc iii}]\,5007 and H$\beta$) and the assigned redshift. The line fluxes were then measured using an {\sc idl} script. Although the comparison of those lines with the H$\alpha$ line fluxes is influenced by many factors (e.g.\ H$\alpha$/[N{\sc ii}] ratio, exact location of the H$\alpha$ line within the filter profile, fraction of emission line light falling into $z$-COSMOS slit) it is noteworthy that both the mean ratio of H$\alpha$/[O{\sc ii}]\,3727\,$=2.27$ and the ratio of H$\alpha$/H$\beta=4.16$  are consistent with an H$\alpha$ extinction of $\sim1$\,mag or slightly higher.

%
%
\begin{figure}
\centering
\includegraphics[width=8.2cm]{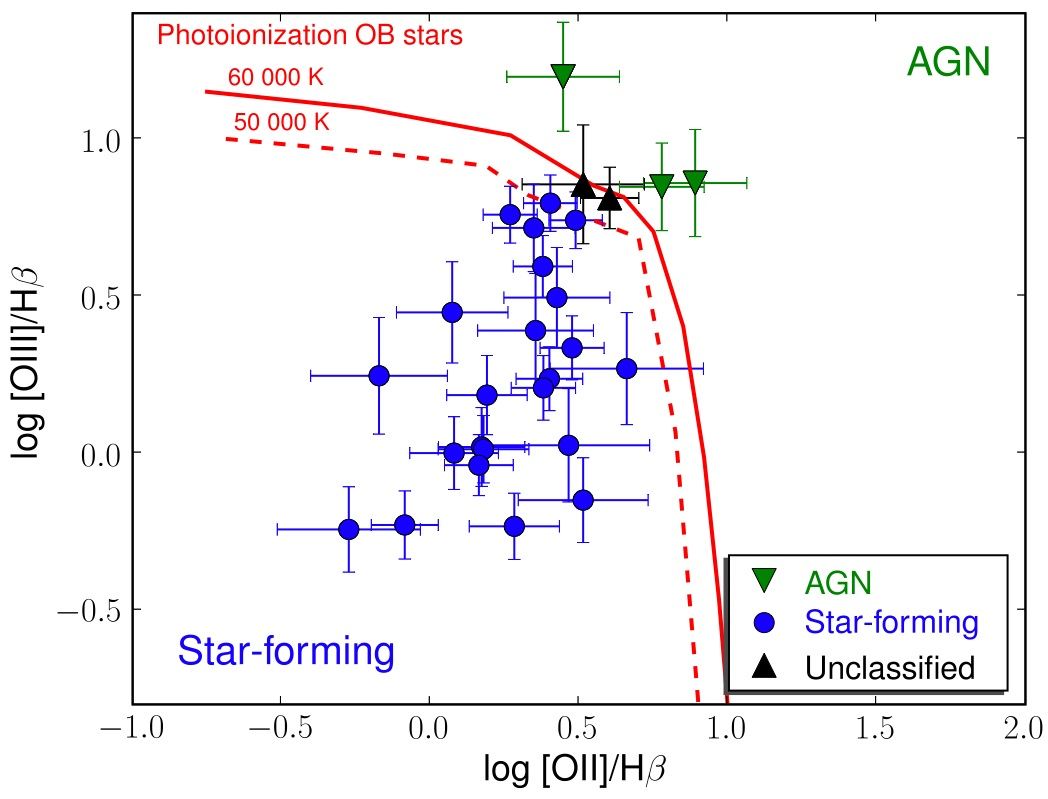}
\caption[AGN contamination]{Line ratios from the $z$-COSMOS spectra of the $z=0.845$ H$\alpha$ sample.  These show that the great majority of the sample is composed of star-forming galaxies (82\%), as expected, with 11\% showing evidence for being AGN contaminants and 7\% being unclassified. The red curves represent the maximum line ratios for a star-forming galaxy (from OB stars with effective temperatures of 60000\,K (solid line) and 50000\,K (dashed line)).  \label{agn}}
\end{figure}

%
%
\begin{figure*}
\begin{minipage}[b]{0.48\linewidth} 
\centering
\includegraphics[width=8.2cm]{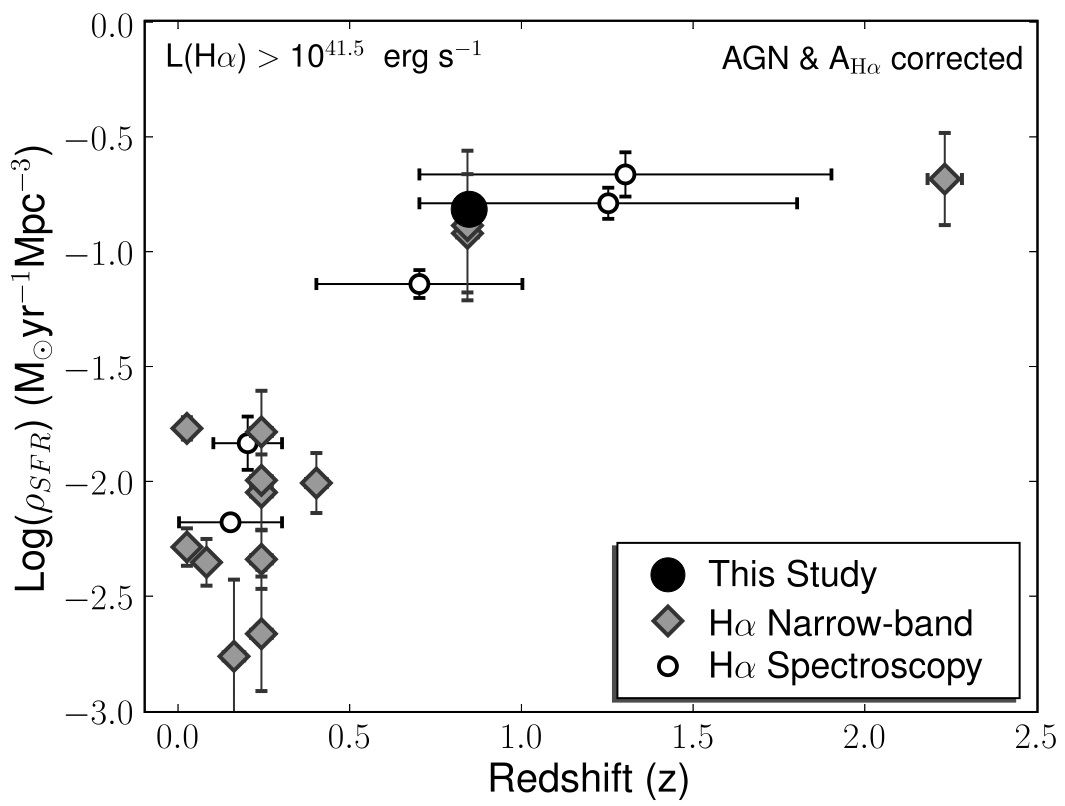}
\end{minipage}
\hspace{0.1cm} 
\begin{minipage}[b]{0.48\linewidth}
\centering
\includegraphics[width=8.2cm]{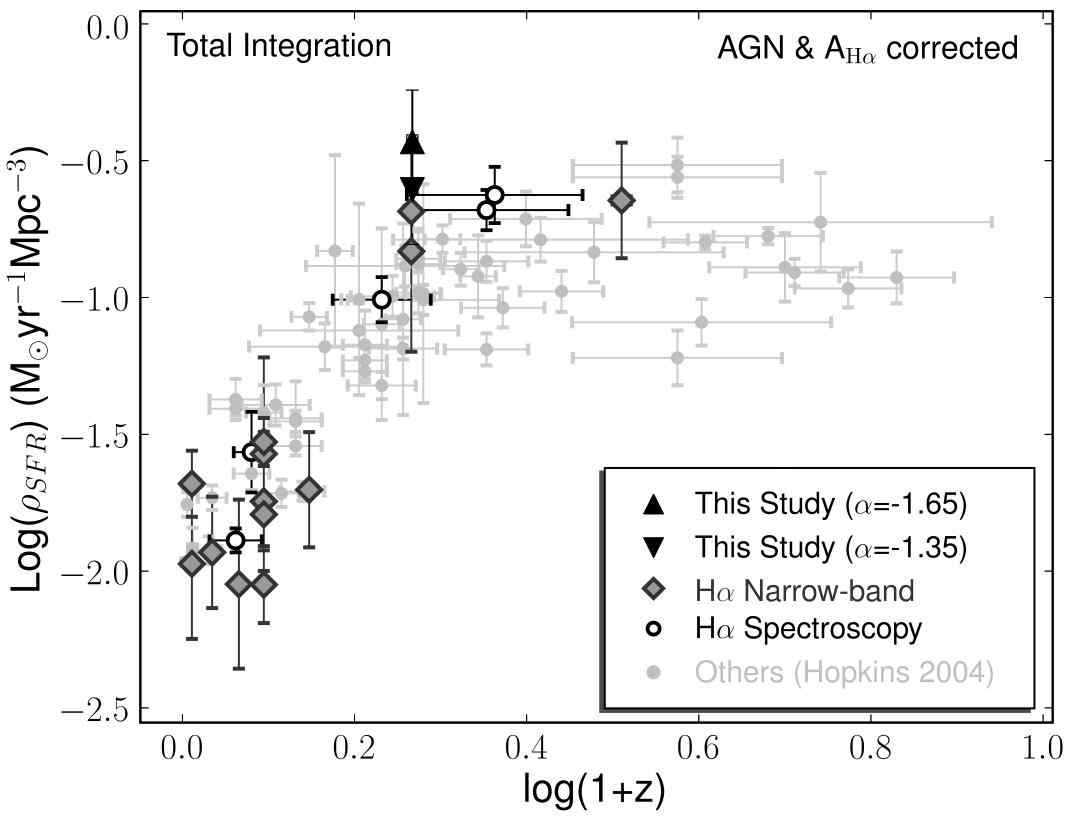}
\end{minipage}
\caption[SFR evolution]{The evolution of the star formation rate density as a function of redshift based on H$\alpha$ -- down to the limit of this survey (left panel) and by integrating the entire luminosity function (right panel). The right panel also includes data from all other star-formation tracers, as detailed in the Hopkins (2004) compilation. Both panels show a clear rise in $\rho_{SFR}$ up to at least $z\sim0.9$--1.0, slightly steeper than the canonical $(1+z)^4$ rise, followed by a flattening out to at least $z\sim 2.2$. H$\alpha$ data points refer to (in order of increasing mean redshift) Gallego et al.\ (1995; $z=0.02$), P{\'e}rez-Gonz{\'a}lez et al.\ (2003; $z=0.02$), Ly et al.\ (2007; $z=0.08$), Sullivan et al.\ (2000; $z=0.15$), Dale et al.\ (2008; $z=0.16$), Tresse \& Maddox (1998; $z=0.20$), Shioya et al.\ (2008; $z=0.24$), Fujita et al.\ (2003; $z=0.24$), Dale et al.\ (2008; $z=0.24$), Morioka et al.\ (2008; $z=0.24$), Ly et al.\ (2007; $z=0.24$), Ly et al.\ (2007; $z=0.4$), Tresse et al.\ (2002; $z=0.7$), Villar et al.\ (2008; $z=0.84$, 1 mag extinction -- higher $\rho_{SFR}$ -- and author's extinction correction), this study ($z=0.845$), Hopkins et al.\ (2000; $z=1.25$), Yan et al.\ (1999; $z=1.3$), Geach et al.\ (2008; $z=2.23$).  \label{sfr_evo}}
\end{figure*}

In order to estimate the AGN contamination, the [O{\sc ii}]\,3727/H$\beta$ and [O{\sc iii}]\,5007/H$\beta$ line ratios were used; these have been widely used to separate AGN from star-forming galaxies \citep[e.g.][]{1997MNRAS.289..419R}. Only spectra with all lines being detected at S/N\,$>3.0$ were used, which results in a sample of 28 galaxies, mainly due to the low S/N at longer wavelengths where [O{\sc iii}]\,5007 and H$\beta$ are found. Figure \ref{agn} shows data-points for the line ratios, while the curves represent maximum line ratios for a star-forming galaxy (from OB stars with effective temperatures of 60000\,K and 50000\,K). From the sample of 28 H$\alpha$ emitters, 23 seem to be clear star-forming galaxies, while 3 are likely to be AGN contaminants. A $\sim15$\% AGN contamination is thus estimated, consistent with that found in other H$\alpha$ studies. The AGN are found to have H$\alpha$ fluxes typical of the rest of the sample.

\subsubsection{Star formation rate density}

The observed H$\alpha$ luminosity function can be used to estimate the average star formation rate density,  $\rho_ {SFR}$, at $z=0.84$. To do this, the standard calibration of Kennicutt (1998) is used to convert the extinction-corrected H$\alpha$ luminosity to a star formation rate:

\begin{equation}
{\rm SFR}({\rm M}_{\odot} {\rm year^{-1}})= 7.9\times 10^{-42} ~{\rm L}_{\rm H\alpha} ~ ({\rm erg s}^{-1}).
\end{equation}
This assumes continuous star formation, Case B recombination at $T_e = 10^4$\,K and a Salpeter initial mass function ranging from 0.1--100\,M$_{\odot}$. All measurements of $\rho_{SFR}$ include a correction of 15\% for AGN contamination and an extinction correction A$_{H_\alpha}=1$\,mag, except where the authors only presented their own extinction corrected luminosity function.

In \S\ref{lf} a significant evolution in the observed H$\alpha$ luminosity function was observed. The left panel of Figure~\ref{sfr_evo} shows how this translates into an evolution in $\rho_{SFR}$ as a function of redshift, for luminosity functions which have been integrated down to L$_{\rm H\alpha}>10^{41.5}$erg\,s$^{-1}$ (the limit of this survey). The measurement at $z=0.84$ presented in this study ($0.15\pm0.02$\,M$_{\odot}$\,yr$^{-1}$\,Mpc$^{-3}$) demonstrates a strong rise in $\rho_{SFR}$, when compared to the local Universe \citep{galo, 2003ApJ...591..827P,2007ApJ...657..738L} and low redshift measurements \citep[e.g.][]{1998ApJ...495..691T, sullivan01,2008AJ....135.1412D,2008arXiv0807.0101M,2008MNRAS.383..339W, 2008ApJS..175..128S,2009arXiv0902.2064S}, as suggested by other smaller surveys done at similar redshifts \citep[e.g.][]{alpha1,2008ApJ...677..169V}. This rise seems to be slightly steeper than $\rho_{SFR}\sim(1+z)^4$. When compared to higher redshift (e.g.\ Geach et al.\ 2008), the observations also support a flattening in $\rho_{SFR}$ around $z\sim 1$, up to at least $z=2.23$. A rise and subsequent flattening of the star formation rate density out to $z\sim2$ has therefore been accurately measured using a single star formation tracer. Cosmic evolution of dust reddening corrections may alter the results slightly but would have to be very strong to change the overall conclusions.

Figure~\ref{sfr_evo} also presents the same evolution, but now integrating the entire luminosity function. Caution should be used in interpreting this figure as it involves extrapolating all the luminosity functions and it is critically dependent on the assumed faint-end slope. For this study, for example, $\rho_{SFR}=0.37\pm0.18$\,M$_{\odot}$\,yr$^{-1}$\,Mpc$^{-3}$ using the measured value of $\alpha=-1.65$, but if one adopts $\alpha=-1.35$, $\rho_{SFR}$ is reduced to $0.28\pm0.08$\,M$_{\odot}$\,yr$^{-1}$\,Mpc$^{-3}$. Measurements obtained from a set of other star-formation indicators compiled by Hopkins (2004) are also shown for comparison (corrected by a common extinction factor consistent with the H$\alpha$ extinction correction applied here). This confirms the same rise seen using only H$\alpha$ and the possible flattening at $z\sim1$. However, it appears the H$\alpha$ measurements at $z>0.7$ are finding slightly higher values of $\rho_{SFR}$ when compared to the average of all other measurements, while at $z<0.3$ they measure slightly lower values. This may reflect a number of systematic errors or factors such as evolution in the typical reddening in the star forming population.

\subsection{The morphology of H$\alpha$ emitters}

\subsubsection{Morphology with {\it HST} imaging} \label{morfo}

\begin{figure*}
\centering
\includegraphics[width=16.6cm]{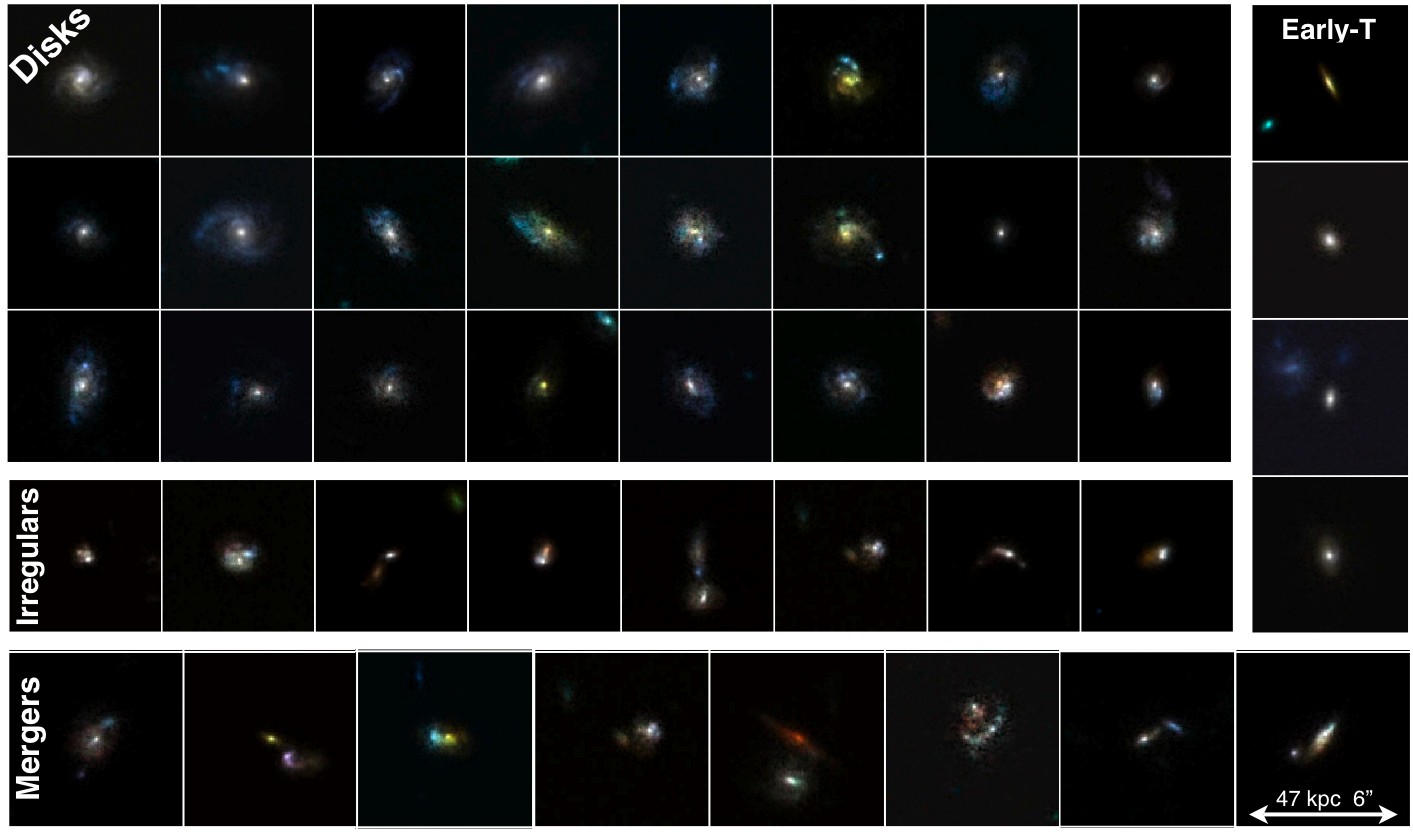}
\caption[Spirals and Mergers in the H$\alpha$ sample]{{\it HST} imaging of examples of the sample of H$\alpha$ emitters at $z=0.84$ in the COSMOS field, organized into the visual morphological classes described in \S\ref{morfo}. These include large and small spirals, mergers and irregulars. The thumbnails are all 6$''\times$\,6$''$ corresponding to 47\,$\times$\,47\,kpc at $z=0.84$. \label{thumbs_int1}}
\end{figure*}

The COSMOS field has sensitive {\it Hubble Space Telescope} ({\it HST}) ACS  F814W coverage which provides detailed morphological information on the H$\alpha$ sample. These data are used to study the morphologies of the selected star-forming galaxies. This morphological information can be greatly enriched with colour information, and thus the ACS images are combined with deep Subaru data in $Brz$ bands to produce a pseudo-true colour image. The Subaru images are first registered and transformed to match the {\it  HST } images  using  {\sc iraf}  and {\sc Python
} scripts. The {\it HST} image is then used to define the luminance of the true colour image. In this way colour and morphological information were derived on $\sim 500$ H$\alpha$ emitters with sufficient detail that a visual morphology analysis could be undertaken. Figure \ref{thumbs_int1} presents some examples of these.

%
%
%
%
\begin{table*}
 \centering
  \caption{Visual morphology classifications compared to the automated classifications obtained by {\sc zest} (Scarlata et al.\ 2007) for the H$\alpha$ emitters at z=0.84. The two agree very well. Differences arise mostly from using colour information for the visual classifications, while {\sc zest} only uses the {\it HST} ${F814W}$ imaging. The final rows of the table give the distribution of visual morphologies for different merger classes.}
  \begin{tabular}{@{}ccccccccc@{}}
  \hline
   Visual & ZEST & ZEST & ZEST & ZEST& Visual & Visual & Visual & Visual \\
   Class & Early & Disks & Irregulars & Unclassified & Total & Non-mergers & Potential Mergers & Mergers \\
 \hline
Early & 7 & 3 & 0 & 4 & 14 (3\%) & 12 & 0 & 2\\
Disks & 2 & 301 & 15 & 63 & 381 (80\%) & 297 & 33 & 51\\
Irregulars & 0 & 10 & 44 & 14 & 68 (14\%) & 1 & 7 & 60 \\
Unclassified & 0 & 2 & 0 & 12  & 14 (3\%) & 0 & 0 & 0 \\
\noalign{\smallskip}
Total & 9 (2\%) & 316 (67\%) & 59 (12\%) & 93 (19\%) & 477 (100\%) & 310 (67\%) & 40 (9\%) & 113 (24\%)  \\
 \hline
\end{tabular}
\label{morphcomp_table}
\end{table*}

All of the galaxies within the sample were visually classified. In order to compare those results with morphologies obtained in an automated way (by {\sc zest}, kindly supplied by Claudia Scarlata, \cite{2007ApJS..172..406S}), 4 morphological classes were used: 1) Early-types, 2) Disks/spirals, 3) Irregulars, and 4) Unclassified. Table \ref{morphcomp_table} shows the comparison between {\sc zest} and the visual classification, while Figure \ref{thumbs_int1} shows examples of each morphological type. In general, a very good agreement between the visual classifications and {\sc zest} was obtained. The main difference is that visual classification was able to reduce the number of unclassified galaxies. Also, visually, it was possible to improve the distinction between disks and irregulars by using the colour information -- distinguishing between multiple bulges and bright spiral arms -- which resulted in a sightly lower fraction of irregulars in the visual classification.

%
%
\begin{figure*}
\begin{minipage}[b]{0.48\linewidth} 
\centering
\includegraphics[width=8.2cm]{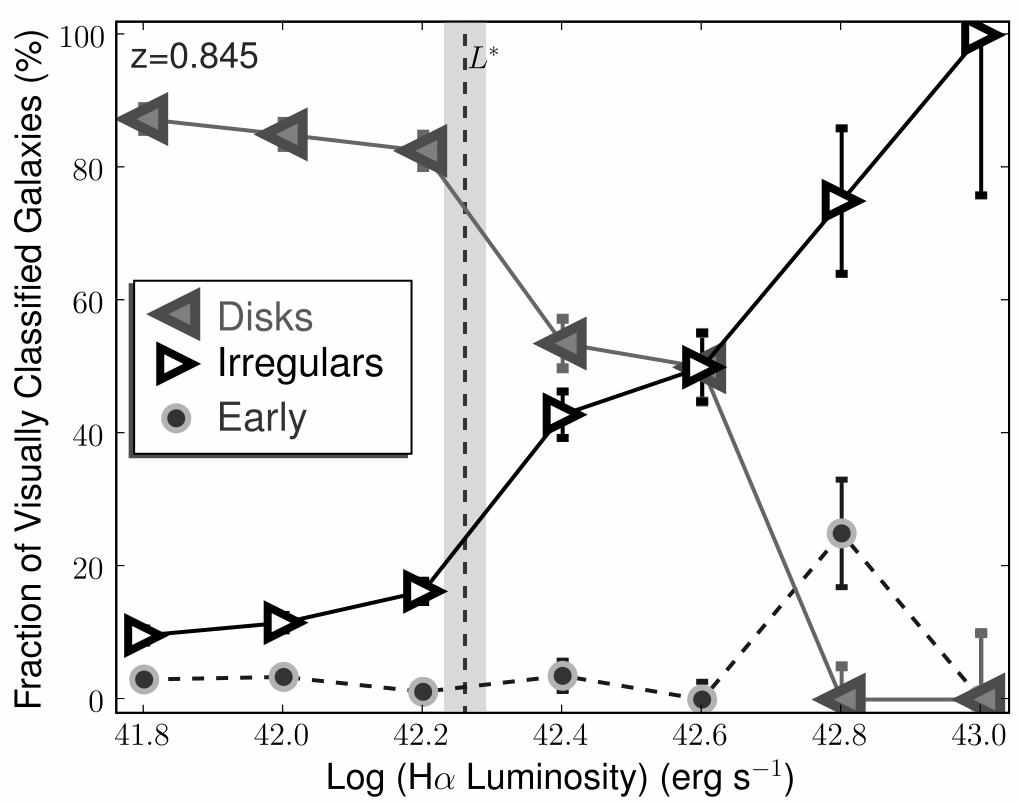}
\end{minipage}
\hspace{0.1cm} 
\begin{minipage}[b]{0.48\linewidth}
\centering
\includegraphics[width=8.2cm,height=6.3cm]{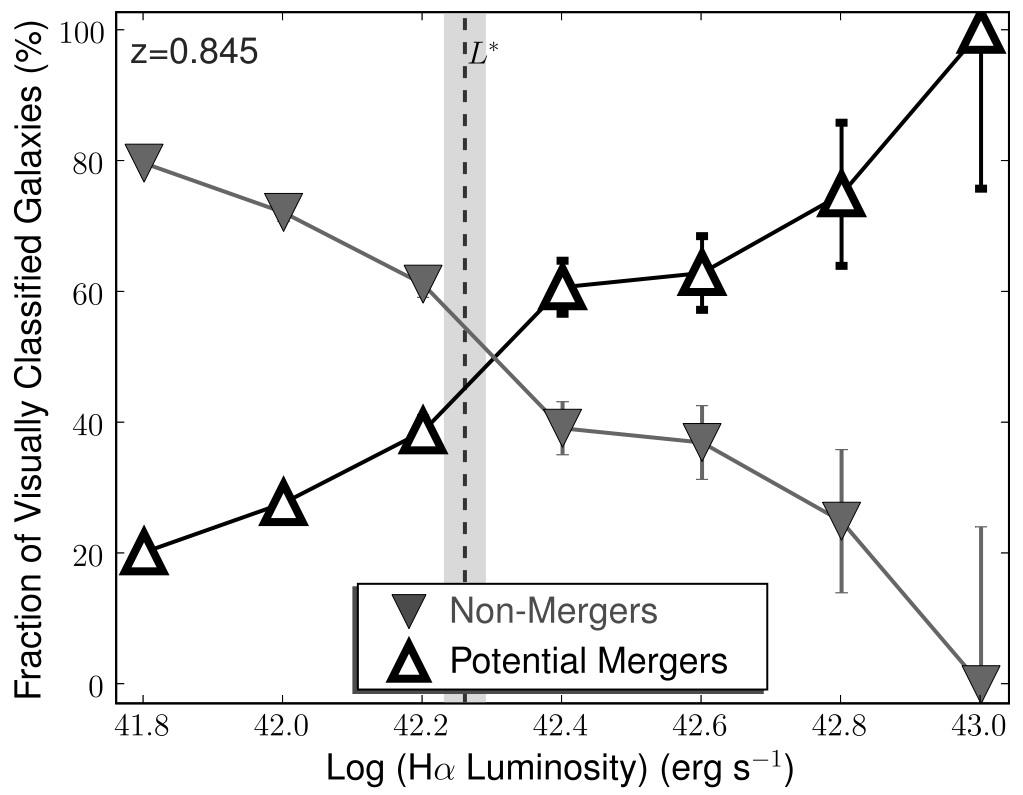}
\end{minipage}
\caption[Spectroscopic-photo]{Morphological class fractions (left panel) and merger fractions (right panel) as a function of H$\alpha$ luminosity at $z=0.84$. These show clear H$\alpha$ luminosity dependences, with the fraction of irregulars and mergers rising with increasing luminosity. The vertical line indicates the value of L$^*$ derived from the best total Schechter function fit with $\alpha=-1.65$. \label{morphdep}}
\end{figure*}

%
%
\begin{figure*}
\begin{minipage}[b]{0.48\linewidth} 
\centering
\includegraphics[width=8.2cm]{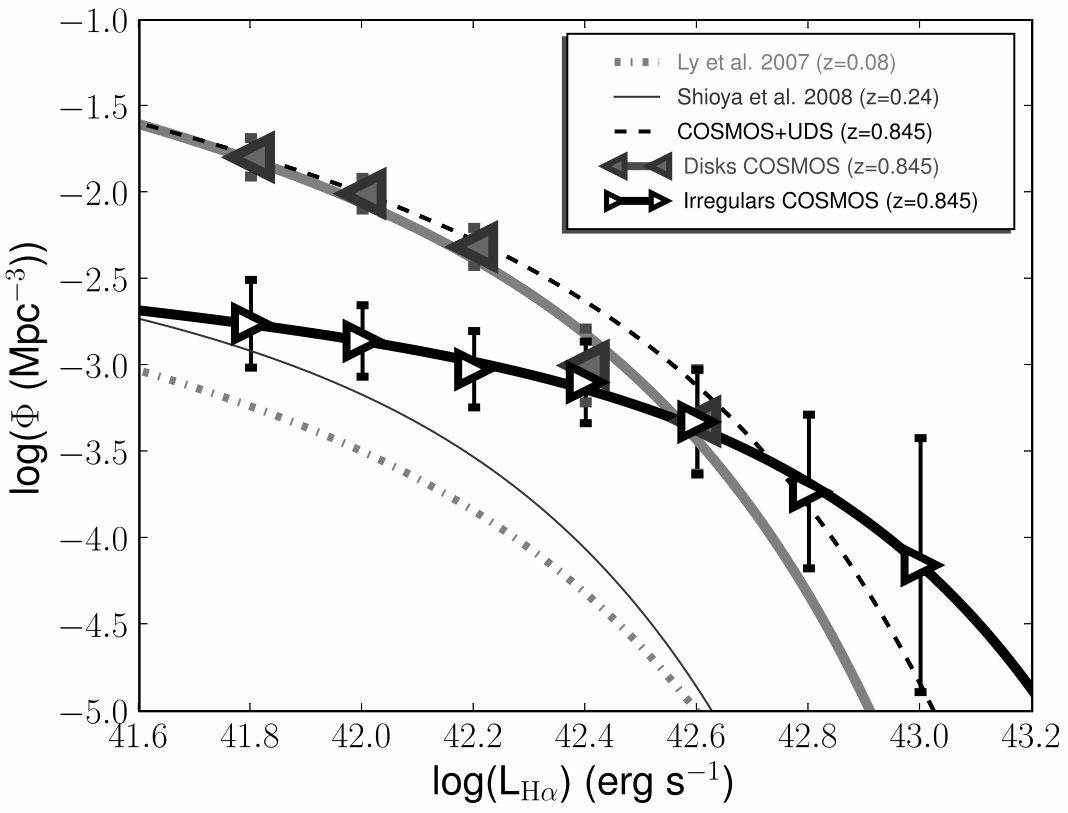}
\end{minipage}
\hspace{0.1cm} 
\begin{minipage}[b]{0.48\linewidth}
\centering
\includegraphics[width=8.2cm,height=6.3cm]{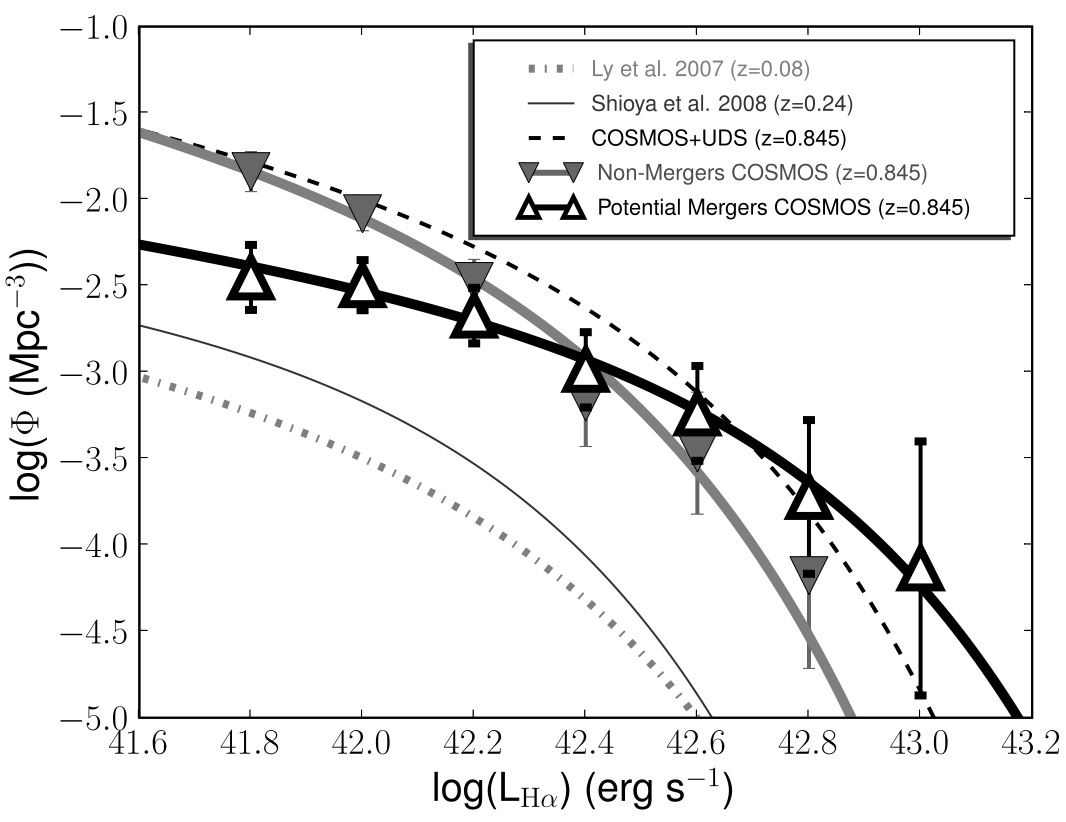}
\end{minipage}
\caption[Spectroscopic-photo]{The computed luminosity functions and the best fits for different visual morphologies. The dash-dotted line refers to the $z=0.08$ LF (Ly et al.\ 2007), with the solid line being that at $z=0.24$ (Shioya et al.\ 2008), and the dashed line is the best fit for the combined UDS+COSMOS luminosity function at $z=0.84$. The brightest bin for both irregulars and potential mergers seems much higher than in the combined luminosity function due to the COSMOS field having a higher space density of bright sources (c.f.\ Figure~5). \label{morphLF}}
\end{figure*}

Due to a considerable number of sources showing evidence of merging activity, the sample was also classified independently into merger classes. A source is classified as a merger when: 1) it presents a clearly disturbed morphology or disturbed disk which is inconsistent with being that of a normal disk galaxy; or 2) the source presents more than 1 bright point-like source and the colour information is inconsistent with one of those being a spiral arm; or 3) there are two or more galaxies which are very close ($<15$\,kpc). By applying these criteria, a visual classification was done, where each source was classified at least twice. In the end, 3 merger classes where used to classify all sources: the consistent non-mergers (non-mergers); those galaxies classified at least once as probable mergers (potential mergers); and galaxies always classified as mergers (mergers). The distribution of mergers within the previous morphological classifications and the total numbers can be found in Table~\ref{morphcomp_table} and examples of classified galaxies can also be found in Figure~\ref{thumbs_int1}. 

From a total of 477 H$\alpha$ emitters, 381 are disks (80\%), with 68 being classified as irregulars (14\%), 14 are early-types (3\%) and a total of 13 (3\%) are unclassified as these are too faint. Furthermore, 24\% of the sample seems to be populated by clear mergers, with the total merger fraction being estimated as $28\pm4$\%; almost all irregulars fall into this class.

\subsubsection{Morphology--H$\alpha$ luminosity relations} \label{morfo}

The left panel of Figure \ref{morphdep} presents the fraction of galaxies classified into each morphology as a function of H$\alpha$ luminosity. There is a clear evolution of morphological type with increasing H$\alpha$ luminosity. While at low star-formation rates, disks dominate the sample completely ($>85$\%), at higher H$\alpha$ luminosities ($L>L^*$) irregulars become more significant,  reaching 100\% in the highest luminosity bin. With the large sample presented in this work it is possible to derive independent luminosity functions for each of those morphological classes. Those can be seen in Figure~\ref{morphLF} with the best fit Schechter function parameters tabulated in Table~5. They illustrate well the different contributions to the total H$\alpha$ luminosity function by disks and irregulars. Irregulars present a remarkably flat luminosity function, only falling at the highest luminosities, while disks demonstrate a steeper faint end and a much lower space density at the bright end.  Overall disks are the dominant contributors to the total $\rho_{SFR}$, with a large number of galaxies producing stars at rates $<10$\,M$_{\odot}$\,yr$^{-1}$.

As mentioned before, almost all of the irregulars show clear evidence of merging activity. In fact, separating non-mergers from likely mergers (mergers, together with potential mergers weighted by 0.5), a very similar behaviour is found (Figure~\ref{morphdep}). While non-mergers dominate at faint luminosities, the contributions from the two populations cross over at $L\sim L^*$, with mergers dominating the bright-end of the luminosity function. The luminosity functions for non-mergers and potential mergers are presented in Figure \ref{morphLF}. This reveals that non-mergers present the typical disk luminosity function found before, but mergers have a steeper luminosity function than the irregulars. This means that while irregulars on their own are only important in the bright end of the LF, mergers seem to play a dominant role at the bright end (accounting for$\sim60$\% of the $\rho_{SFR}$ there) together with a non-negligible contribution even at the faintest luminosities. Overall, mergers account for $\sim20$\% of the total $\rho_{SFR}$ at $z=0.84$.
 
The relations which have been presented were found to be very robust, with the same results being obtained regardless of the use of the visual or {\sc zest} classifications, and independently of the [N{\sc ii}] correction -- a constant correction also produces the same dependences.

\subsubsection{Redshift evolution of the morphology relations} \label{morfo}

One can compare the morphological mix for the H$\alpha$ sample at $z=0.84$ with that found at lower redshifts.
In the local Universe, the morphologies of the H$\alpha$ emitters seem to be somewhat different, with visual morphologies from \cite{1996A&AS..118....7V} indicating that disks are 88\% of the sample, while early-type galaxies increase their importance (9\%) and irregulars drop to only 3\%. At $z=0.24$ the narrow-band H$\alpha$ emitters of Shioya et al.\ (2008) in the COSMOS field can be more directly compared to the $z=0.84$ sample, using {\sc zest} morphologies for both (although noting that the rest-frame wavelength of the images is different between the two studies -- rest-frame $B$ band at $z=0.84$ and rest-frame $R$ band at $z=0.24$ -- which could introduce a small bias). In the  {\sc zest}  classifications at $z=0.24$, disks are still dominant (89\%), $\sim$10\% of the galaxies have an irregular morphology, and elliptical galaxies account for only 1\% of the sample. These results seem to show that while disks are always the dominant population ($>80$\%) up to $z\sim1$, there is a significant increase in the irregular fraction from 3\% at $z=0$ to 10\% at $z=0.24$ and 15\% at $z=0.84$.

%
%
\begin{figure}
\centering
\includegraphics[width=8.2cm]{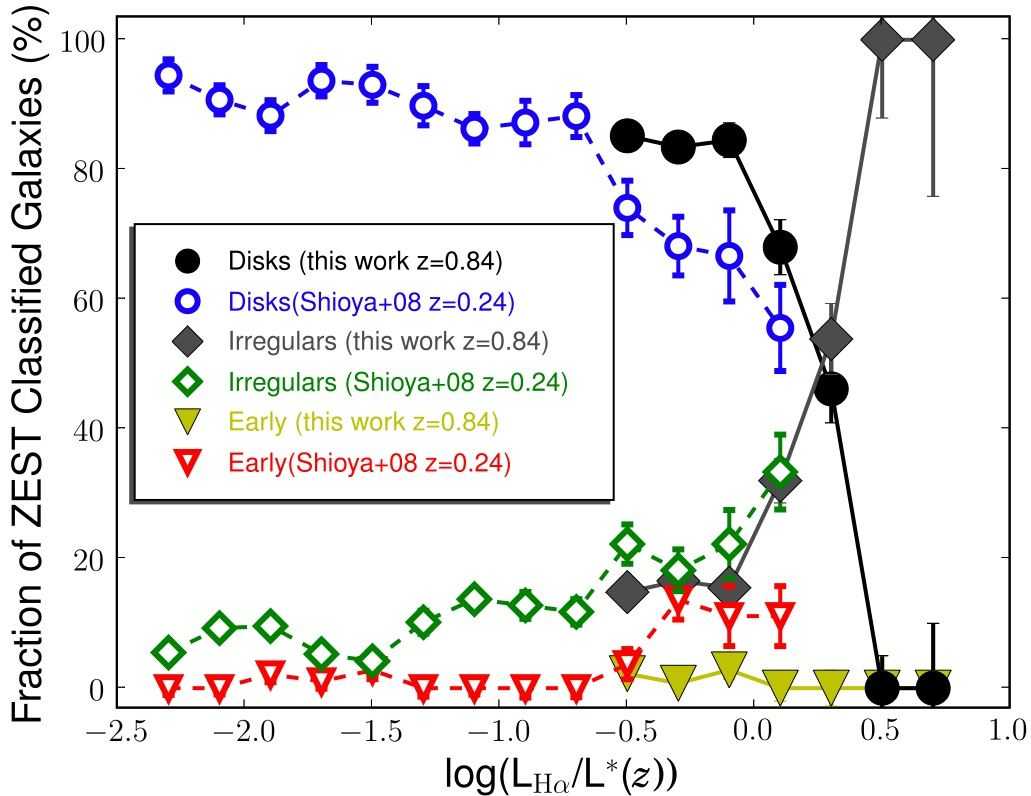}
\caption[Spectroscopic-photo]{Morphology class fractions as a function of H$\alpha$ luminosity for $z=0.24$ (Shioya et al.\ 2008) and this study. These results find the same morphological dependence at both epochs consistent with an evolution in $L^*$ ($\log(L)\sim0.4$). Morphologies were obtained in an automated way by {\sc zest} (Scarlata et al.\ 2007) for both samples.  \label{morphdepzlow}}
\end{figure}

\begin{table*}
 \centering
  \caption{The H$\alpha$ Luminosity function and $\rho_{SFR}$ for different subsets of the data. Both log$\rho_{L \ >41.5}$ and $\rho_{SFR \ >41.5}$ result from integrating the H$\alpha$ luminosity function down to $10^{41.5}$\,erg\,s$^{-1}$. $\rho_{SFR \ total}$ results from a complete integration of the luminosity function. A 15\% AGN correction is made to all $\rho_{SFR}$ measurements.}
  \begin{tabular}{@{}cccccccc@{}}
  \hline
   Sample & log$\phi^*$ & log$L^*$& $\alpha$ & log$\rho_{L \ >41.5}$ & $\rho_{SFR \ >41.5}$ & $\rho_{SFR \ total}$\\
      \noalign{\smallskip}
     & (Mpc$^{-3}$) & (erg\,s$^{-1}$) & & (erg\,s$^{-1}$Mpc$^{-3}$) & (M$_{\odot}$\,yr$^{-1}$\,Mpc$^{-3}$) & (M$_{\odot}$\,yr$^{-1}$\,Mpc$^{-3}$)\\

 \hline
Total: COSMOS+UDS & $-$1.92$\pm$0.10 & 42.26$\pm$0.05 & $-$1.65$\pm$0.15 &  40.37$\pm$0.04 & 0.15$\pm$0.01  &  0.37$\pm$0.15  \\
UDS & $-$1.98$\pm$0.16 & 42.25$\pm$0.10 & $-$1.70$\pm$0.22 &  40.32$\pm$0.06 & 0.14$\pm$0.03  &  0.39$\pm$0.18  \\
COSMOS & $-$1.90$\pm$0.12 & 42.28$\pm$0.07 & $-$1.68$\pm$0.18 &  40.44$\pm$0.05 & 0.18$\pm$0.03  &  0.45$\pm$0.22  \\
 \noalign{\smallskip}
  \hline
COSMOS: Disks & $-$1.81$\pm$0.13 & 42.12$\pm$0.06 & $-$1.65$\pm$0.22 & 40.27$\pm$0.05 & 0.13$\pm$0.02 & 0.41$\pm$0.18  \\
COSMOS: Irregulars & $-$2.90$\pm$0.23 & 42.58$\pm$0.14 & $-$1.27$\pm$0.21 & 39.70$\pm$0.10  & 0.03$\pm$0.01  & 0.04$\pm$0.01 \\
 \noalign{\smallskip}
  \hline
COSMOS: Non-mergers & $-$1.96$\pm$0.14 & 42.16$\pm$0.09 & $-$1.71$\pm$0.21 &  40.19$\pm$0.06 & 0.10$\pm$0.02 & 0.33$\pm$0.20 \\
COSMOS: Potential Mergers & $-$2.63$\pm$0.13 & 42.50$\pm$0.10 & $-$1.47$\pm$0.20 &  39.94$\pm$0.09 & 0.06$\pm$0.01 & 0.08$\pm$0.04  \\
 \hline
\end{tabular}
\label{lf_morph}
\end{table*}

The {\sc zest} classifications for the Shioya et al.\ (2008) sample can also be compared with those for the $z=0.84$ sample to see if there is any evolution in the morphology--H$\alpha$ luminosity relations.  Thus
Figure~\ref{morphdepzlow} presents the comparison between $z=0.24$ and $z=0.84$ with the luminosities scaled by $L^*$. A very similar variation in morphological mix with H$\alpha$ luminosity  is found at $z=0.24$, with the irregular fraction increasing from $\sim10$\% to $\sim35$\% with increasing luminosity and with the disk fraction having the opposite behaviour, decreasing from  $\sim 90$\% to $\sim60$\%. The relatively good agreement between the data at these two different redshifts and the fact that the switch-over luminosity appears to be at the same $L/L^*$ at both redshifts seems to point towards a rather simple $L^*$ evolution driving the morphology--H$\alpha$ luminosity relation. However, the Shioya et al.\ (2008) sample is not able to probe the brightest sources and thus there is no low redshift data to directly compare the complete dominance of irregulars at $L>3 L^*$.

The disk fraction at $z=0.24$ appears to be slightly lower than the disk fraction at $z=0.84$ (for the bins which can be directly compared). Curiously, this is due to an apparent rise in the fraction of early-types.  However, this might be simply reflecting classification errors between the two epochs. At least some of the very faint galaxies which have been unclassified at $z=0.84$ (even after visual classification) are likely to be early-types which are just too faint to be seen at $z=0.84$, but sufficiently bright at $z=0.24$ to be classified.

\section{Conclusions}

Deep near-infrared narrow-band imaging has been obtained, allowing the selection of line emitting galaxies down to an effective flux limit of F$_{{\rm H}\alpha}\sim10^{-16}$\,erg\,s$^{-1}$\,cm$^{-2}$.  This has resulted in the largest and deepest survey of emission line selected star forming galaxies at $z\sim1$, detecting 1517 potential line emitters over an area of $\sim 1.4\deg^2$ in the COSMOS and UDS fields, with 1370 having quality multi-wavelength data available in a region of $1.3\deg^2$.  For H$\alpha$ emission line objects this survey probes a co-moving volume of $\sim1.8\times10^5$\,Mpc$^3$ at
$z=0.84$ down to a star formation rate of $\sim3$\,M$_{\odot}$\,yr$^{-1}$ (with an $A_{H\alpha}=1$\,mag extinction correction). 

Photometric redshifts for COSMOS and UKIDSS UDS (Mobasher et al.\ 2007; Cirasuolo et al.\ 2008) clearly show that the majority of the selected emitters are  H$\alpha$ 
emitters at $z\sim0.85$ with a secondary population of [O{\sc iii}]\,5007/H$\beta$ emitters at $z\sim1.4$--1.5. Almost 120 emitters were confirmed spectroscopically, from which 93 are H$\alpha$ at $z=0.84$. The contamination within the sample of emitters is estimated to be lower than $\sim 6$\%, and the contamination within the H$\alpha$ sample is much lower ($\sim 0$\%) based on the current samples.
A total of 743 H$\alpha$ selected emitters (based on their photometric and spectroscopic redshifts) was obtained in the two fields. These were used to calculate the luminosity function after correcting for [N{\sc ii}] flux contamination, extinction, incompleteness and filter profile biases. The morphologies of these emitters were also investigated. The main conclusions of this work are:

\begin{itemize}

\item The H$\alpha$ luminosity function at $z=0.84$ found is well fitted by a Schechter function with $\phi^*=10^{-1.92\pm0.10}$\,Mpc$^{-3}$, $\alpha=-1.65\pm0.15$ and $L^*=10^{42.26\pm0.05}$\,erg\,s$^{-1}$. This demonstrates a strong evolution in the H$\alpha$ luminosity function compared to lower
redshifts and 
agrees reasonably well with previous smaller studies at $z\sim 1$. 

\item The evolution of the H$\alpha$ luminosity function can be described by an increase in $\phi^*$ and $L^*$, at least out to $z\sim1$, with L$^*$ then continuing to rise up to $z\sim2$ but $\phi^*$ peaking around $z\sim1$ and then decreasing at higher redshifts.

\item The integrated luminosity function is used to estimate the cosmic star formation rate density, ($\rho_{SFR}$) at $z=0.84$: $0.15\pm0.01$\,M$_{\odot}$\,yr$^{-1}$\,Mpc$^{-3}$ (corrected for 15\% AGN contamination and integrated down to 2.5\,M$_{\odot}$\,yr$^{-1}$).

\item An accurate determination of the cosmic evolution of $\rho_{SFR}$ has been made using a single star formation tracer (H$\alpha$) from $z=0$ to $z=2.23$. This shows a strong rise up to $z\sim1$ followed by a flattening out to $z\sim2.2$.

\item H$\alpha$ emitters at $z=0.84$ are mostly morphologically classed as disks in the rest-frame $B$-band ($82\pm3$\%). Irregulars account for $15\pm2$\% of the sample and early-type galaxies are only $3\pm1$\%. Apparent mergers are a significant fraction of the sample ($28\pm4$\%).

\item A strong morphology--H$\alpha$ luminosity relation is found at $z$=$0.84$, with the fraction of irregulars rising steadily with luminosity and the fraction of disks falling. Mergers/non-mergers present the same behaviour, and $L^*$ (from the total sample) seems to define a critical switch-over luminosity between the two populations. 

\item Mergers dominate the bright end of the total H$\alpha$ luminosity function at $z=0.84$ and $\sim20$\% of the total $\rho_{SFR}$ is due to their activity.

\item A similar morphology--H$\alpha$ luminosity relation is found at lower redshift ($z=0.24$), consistent with a simple L$^*$ evolution.

\end{itemize}

These results suggest that the evolution of both the H$\alpha$ luminosity function and $\rho_{SFR}$ change significantly in nature beyond $z\sim1$, and that it is entirely plausible that this is driven by the different evolutionary behaviour of two different populations of star forming galaxies. Out to $z\sim1$, the integrated $\rho_{SFR}$ at each redshift is produced predominantly by disk galaxies; it is therefore the evolution of these disk galaxies, rather than that of major mergers, which drives the strong decrease in the cosmic star formation rate density from redshift one to the current epoch -- this is in line with other recent results \cite[e.g.][]{2008ApJ...672..177L}. At these redshifts, the evolution in $\rho_{SFR}$ arises predominantly from an evolution in the characteristic space density of the H$\alpha$ luminosity function ($\phi^*$; which evolves by more than an order of magnitude between $z=0$ and $z=0.84$), rather than a strong evolution in $L^*$. With their relatively quiescent star formation activity, the disk galaxies dominate the H$\alpha$ luminosity function at low luminosities, and are thus responsible both for the $\phi^*$ evolution and for setting the faint-end slope of the luminosity function.

In contrast, irregular and merging galaxies appear to dominate the H$\alpha$ luminosity function above $L^*$, at $z=0.84$ (and also at $z=0.24$), being responsible for more than 50\% of the non-extrapolated $\rho_{SFR}$ (integrated down to 2.5\,M$_{\odot}$\,yr$^{-1}$). The evolution of these systems controls the bright end of luminosity function, and thus the cosmic evolution of $L^*$. The continued strong evolution of $L^*$ between $z=0.84$ and $z =2.23$, and the decrease in $\phi^*$, suggests an increasing importance of merger-driven star formation activity beyond $z\sim1$, especially as the irregular H$\alpha$ luminosity function seems to be very similar to the one found at $z=2.23$ by Geach et al. (2008) when taking into account an $L^*$ evolution. This is consistent with recent results such as \cite{2009arXiv0903.3035S}. The completed HiZELS survey will provide statistically significant samples of H$\alpha$ emitters at $z=1.47$ and $z=2.23$ and so provide a direct test of this suggestion.

\section*{Acknowledgments}

The authors thank the referee for relevant and detailed suggestions that have improved this work. DS would like to thank FCT for financial support (grant: SFRH/BD/36628/2007). PNB \& IRS acknowledge the Royal Society. JEG thanks the U.K. Science and Technology Facility Council (STFC) and KC acknowledges for a STFC Fellowship. The authors would also like to thank Claudia Scarlata for kindly supplying {\sc zest} morphologies for the COSMOS samples, the ESO Large Program 175.A-0839 ($z$-COSMOS), Jesus Gallego and Pablo P{\'e}rez-Gonz{\'a}lez for helpful discussions,  and Andy Adamson, Luca Rizzi and Tim, Thor and Jack for the support on the UKIRT telescope.

\bibliographystyle{mn2e.bst}
\bibliography{bibliography.bib}

\appendix

\bsp

\label{lastpage}

\end{document}